\documentclass[format=acmsmall, review=false, screen=true]{acmart}

\usepackage{booktabs} 

\usepackage[utf8]{inputenc}
\usepackage{graphicx}
\usepackage{ulem}
\usepackage{todonotes}

\usepackage{blindtext}
\usepackage{acro}
\acsetup{first-style=reversed}
\usepackage[linesnumbered,boxed,vlined]{algorithm2e}

\usepackage[english]{babel}
\usepackage{wrapfig}
\usepackage{multirow}
\usepackage{listings}
\usepackage{url}
\usepackage{booktabs}
\usepackage{array}
\usepackage{hyperref}
\usepackage[font=small, labelfont=bf]{caption}
\usepackage{subcaption}
\usepackage{makecell, booktabs}
\usepackage[flushleft]{threeparttable}
\usepackage{comment}
\usepackage{tabularx}

\SetAlCapFnt{\small} 
\SetAlCapNameFnt{\small} 
\SetAlCapSkip{.51em}

\DeclareMathSizes{10}{8}{8}{6}

\newcommand\tab[1][1cm]{\hspace*{#1}}

\acmVolume{0}
\acmNumber{0}
\acmArticle{39}
\acmYear{2018}
\acmMonth{6}
\copyrightyear{2018}

\setcopyright{acmlicensed}

\acmDOI{0000001.0000001}

\received{June 2018}

\DeclareAcronym{IDS}{short=IDS,long=Intrusion Detection System}  
\DeclareAcronym{NIDS}{short=NIDS,long=Network Intrusion Detection System}
\DeclareAcronym{ML}{short=ML,long=Machine Learning}
\DeclareAcronym{ID2T}{short=ID2T,long=Intrusion Detection Dataset Toolkit}
\DeclareAcronym{PCAP}{short=PCAP,long=Packet Capture}
\DeclareAcronym{TTL}{short=TTL,long=Time To Live}
\DeclareAcronym{API}{short=API,long=Application Programming Interface}
\DeclareAcronym{DDoS}{short=DDoS,long=Distributed Denial of Service}
\DeclareAcronym{DoS}{short=DoS,long=Denial of Service}
\DeclareAcronym{TIDED}{short=TIDED,long=Testing Intrusion Detection Datasets}
\DeclareAcronym{CAIDA}{short=CAIDA,long=Cooperative Association for Internet Data Analysis}
\DeclareAcronym{DARPA}{short=DARPA,long=Defense Advanced Research Projects Agency}
\DeclareAcronym{MIT}{short=MIT,long=Massachusetts Institute of Technology}
\DeclareAcronym{CDX}{short=CDX,long=Cyber Defense Exercise}
\DeclareAcronym{NSA}{short=NSA,long=National Security Agency}
\DeclareAcronym{USMA}{short=USMA,long=U.S Military Academy}
\DeclareAcronym{DNS}{short=DNS,long=Domain Name System}
\DeclareAcronym{LBNL}{short=LBNL,long=Lawrence Berkeley National Laboratory}
\DeclareAcronym{KDD}{short=KDD,long=Knowledge Discovery and Data Mining}
\DeclareAcronym{SIGKDD}{short=SIGKDD,long=Special Interest Group on Knowledge Discovery and Data Mining}
\DeclareAcronym{MAWI}{short=MAWI,long=Measurement and Analysis on the WIDE Internet}
\DeclareAcronym{R2L}{short=R2L,long=Remote to Local}
\DeclareAcronym{U2R}{short=U2R,long=User to Root}
\DeclareAcronym{WIDE}{short=WIDE,long=Widely Integrated Distributed Environment}
\DeclareAcronym{IMPACT}{short=IMPACT,long=Information Marketplace for Policy and Analysis of Cyber-risk \& Trust}
\DeclareAcronym{UMass}{short=UMass,long=University of MAssachusetts Amherst}
\DeclareAcronym{ADFALD}{short=ADFALD,long=Australian Defence Force Academy Linux Dataset}
\DeclareAcronym{IRSC}{short=IRSC,long=Indian River State College}
\DeclareAcronym{UNSWNB}{short=UNSWNB,long=University of New South Wale Network Based}
\DeclareAcronym{FLAME}{short=FLAME,long=Flow-Level Anomaly Modeling Engine}
\DeclareAcronym{IRC}{short=IRC,long=Internet Relay Chat}
\DeclareAcronym{MTU}{short=MTU,long=Maximum Transmission Unit}
\DeclareAcronym{RFC}{short=RFC,long=Request for Comments}
\DeclareAcronym{ToS}{short=ToS,long=Type of Service}
\DeclareAcronym{NETAD}{short=NETAD,long=Network Traffic Anomaly Detector}
\DeclareAcronym{IANA}{short=IANA,long=Internet Assigned Numbers Authority}
\DeclareAcronym{MSS}{short=MSS,long=Maximum Segment Size}
\DeclareAcronym{SMB}{short=SMB,long=Server Message Block}
\DeclareAcronym{CSV}{short=CSV,long=Comma Separated Values}
\DeclareAcronym{NBT}{short=NBT,long=NetBIOS over TCP}
\DeclareAcronym{RNN}{short=RNN,long=Replicator Neural Network}
\DeclareAcronym{ICS}{short=ICS,long=Industrial Control Systems}

\begin{document}
\title[On generating network traffic datasets with synthetic attacks for intrusion detection]{On generating network traffic datasets with synthetic attacks for intrusion detection}  

\author{Carlos Garcia Cordero}
\affiliation{%
	\institution{Technische Universität Darmstadt}
	\department{Telecooperation Group}
	\city{Darmstadt}
	\state{Hessen}
	\postcode{64289}
	\country{Germany}
}
\author{Emmanouil Vasilomanolakis}
\affiliation{%
	\institution{Aalborg University}
	\department{Electronic Systems, Center for Communication, Media and Information technologies}
	\city{Copenhagen}
	\postcode{2450}
	\country{Denmark}
}
\author{Aidmar Wainakh}
\author{Max Mühlhäuser}
\affiliation{%
  \institution{Technische Universität Darmstadt}
  \department{Telecooperation Group}
  \city{Darmstadt}
  \state{Hessen}
  \postcode{64289}
  \country{Germany}
}

\author{Simin Nadjm-Tehrani}
\affiliation{%
	\institution{Linköping University}
	\department{Real-time Systems Laboratory}
	\city{Linköping}
	\postcode{S-581 83}
	\country{Sweden}
}

\begin{abstract}
Most research in the area of intrusion detection requires datasets to develop, evaluate or compare systems in one way or another. In this field, however, finding suitable datasets is a challenge on to itself. Most publicly available datasets have negative qualities that limit their usefulness. In this article, we propose \Ac{ID2T} to tackle this problem. \Ac{ID2T} facilitates the creation of labeled datasets by injecting synthetic attacks into background traffic. The injected synthetic attacks blend themselves with the background traffic by mimicking the background traffic's properties to eliminate any trace of \Ac{ID2T}'s usage.

This work has three core contribution areas. First, we present a comprehensive survey on intrusion detection datasets. In the survey, we propose a classification to group the negative qualities we found in the datasets. Second, the architecture of \Ac{ID2T} is revised, improved and expanded. The architectural changes enable \Ac{ID2T} to inject recent and advanced attacks such as the widespread EternalBlue exploit or botnet communication patterns. The toolkit's new functionality provides a set of tests, known as \Ac{TIDED}, that help identify potential defects in the background traffic into which attacks are injected. Third, we illustrate how \Ac{ID2T} is used in different use-case scenarios to evaluate the performance of anomaly and signature-based intrusion detection systems in a reproducible manner. \Ac{ID2T} is open source software and is made available to the community to expand its arsenal of attacks and capabilities.


\end{abstract}

%
%

%
%


\maketitle

\renewcommand{\shortauthors}{Carlos Garcia C. et al.}

\section{Introduction}\label{sec:introduction}

Evaluating the detection capabilities of \acp{NIDS} has become a crucial task in today's Internet age \cite{vasilomanolakis2015survey}. It is not only due to our dependency on the Internet, but also because of the Internet's threat landscape that \acp{NIDS} have become an almost mandatory line of defense against attacks. This need to develop \acp{NIDS} that can keep up with evolving attacks and motivated adversaries has yielded much research in the direction of identifying old and new, previously unobserved, threats. Evaluating \ac{NIDS} has intrinsic complexities and challenges that need to be addressed irrespective of which intrusion detection method they use. The evaluation of \acp{NIDS} relies on quality datasets to assess their capabilities. Quality datasets are also what enables accurate comparison of different intrusion detection methods. Reliable datasets, however, are not readily available.

Reliable datasets useful for the evaluation of \acp{NIDS} are hard to obtain. Widely available datasets tend to be outdated, often lack labeled attacks and usually contain overlooked defects. Furthermore, these datasets are often not publicly available or difficult to obtain \cite{kock2014datasets}. Most datasets are distributed as \ac{PCAP} files. These files, being in essence just an ordered collection of network packets, can be thought as of having packets originating from one of two sources: a benign or a malicious one. Packets originating from benign sources compose the \emph{normal} or \emph{background} traffic of the dataset. Packets created by malicious activities on a network compose the \emph{attack} traffic of the dataset. To overcome the seemingly inherent difficulties of creating and sharing datasets, we have developed \ac{ID2T}. \ac{ID2T} is a tool that creates and injects synthetic attack traffic into background traffic. Rather that just na\"ively merging together the background and attack traffic, \ac{ID2T} attempts to replicate the properties of the background traffic in the synthetic attack traffic. The level by which properties are replicated is controlled by the user to suite their needs.

Users of \ac{ID2T} are not confined to only injecting the set of attacks provided. The instruments to develop new attacks are provided. These come in the form of an \ac{API} to easily extract, compare and replicate traffic properties.

\begin{figure}[ht]
  \centering
  \includegraphics[width=0.65\textwidth]{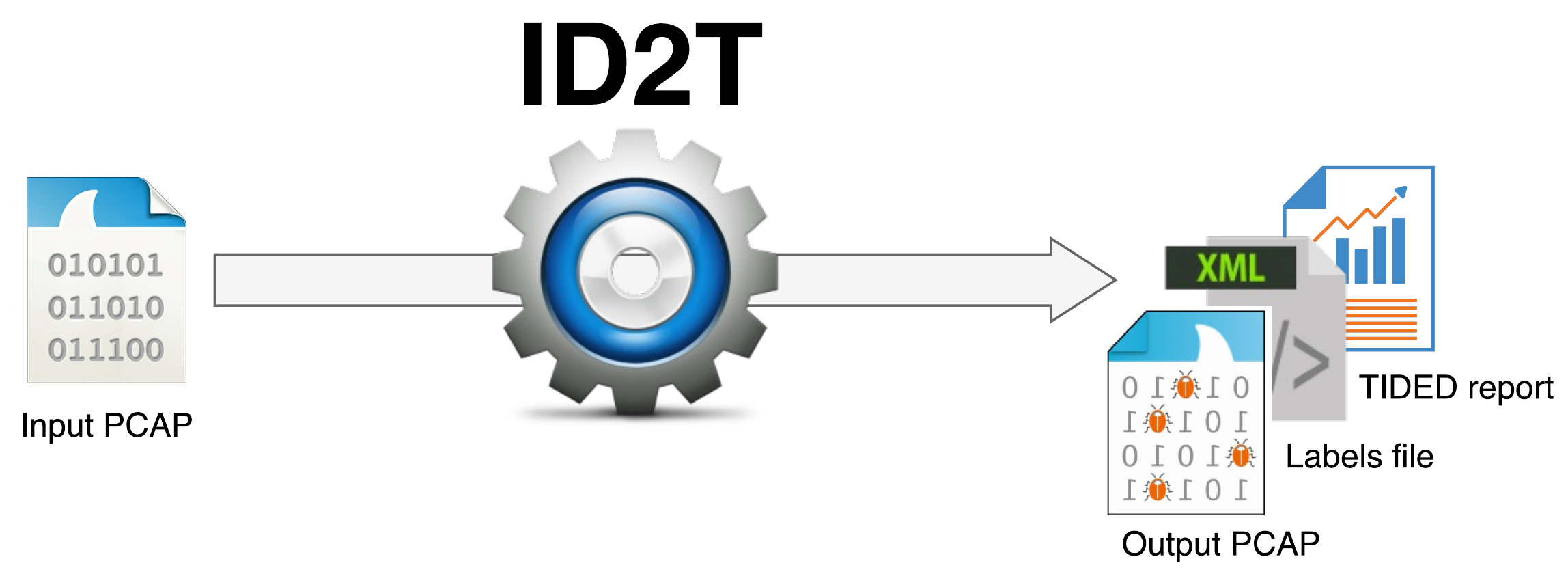}
  \caption{ID2T provides all what is needed to create datasets that contain attacks that replicate the properties of arbitrary background traffic.  The output dataset, in the form of a PCAP file, contains labeled attacks that suite the needs of the user.}
  \label{fig:id2t-gears}
\end{figure}

The design philosophy of \ac{ID2T} is simple: background traffic is provided by the user and \ac{ID2T} adds attacks to the background traffic following the specifications of the user. The injected attacks are labeled and clearly identified for intrusion detection mechanisms to use; for instance, for evaluation purposes. Figure \ref{fig:id2t-gears} shows the main architectural components that together interact to create datasets with synthetic attacks that replicate background traffic's properties. The inputs of \ac{ID2T} are a \ac{PCAP} file containing background data and user  supplied parameters. Its output is a \ac{PCAP} file injected with synthetic attacks, a labels file specifying the time and location of the injected attacks, and a report on the statistical analysis of the background traffic.

\subsection{Contributions}

This work substantially extends two earlier publications \cite{Cordero2015id2t, vasilomanolakis2016id2t}. The first publication presents the idea of \Ac{ID2T} in the form of a poster~\cite{Cordero2015id2t}. In the poster, we presented a basic architecture for injecting attacks that replicate background traffic's properties along with the basic requirements to do so. In the second work, we formalized the requirements for injecting attacks, we built a prototype, and we evaluated the prototype's performance and ability to inject attacks~\cite{vasilomanolakis2016id2t}. This work has the following additional contributions:
\begin{itemize}
  \item We present a comprehensive survey of \Ac{NIDS} datasets and synthetic dataset generation tools as well as a categorization of the problems found in these.
  \item We release \Ac{ID2T} as an open source\footnote{The source code of ID2T can be found in https://github.com/tklab-tud/ID2T.} software to help \Ac{NIDS} researchers to create, distribute and replicate datasets.
  \item We develop the \ac{ID2T} module named \ac{TIDED} that calculates quantitative characteristics of network traffic that help researchers determine whether background network traffic has abnormal characteristics.
  \item We implement a number of recent attacks. For instance, these include the popular Wannacry attack, based on the  \emph{EternalBlue} exploit, as an injectable attack within \Ac{ID2T}. This attack highlights how synthetic attacks can be easily created in contrast to generating and publishing whole new datasets that contain the same attack. We further demonstrate use-case scenarios that show how \Ac{ID2T} can be used to evaluate \Acp{NIDS} in a reproducible way.
\end{itemize}

\subsection{The Limitations of \ac{ID2T}}\label{sec:limit-gener-synth}

\ac{ID2T} limits itself to replicating the properties of user-supplied background traffic into synthetically generated attack traffic. Many attack scenarios are not affected by this limitation, others, however, may be negatively affected. If an attack is not expected to alter the state of a network, the replication strategy employed by \ac{ID2T} is sufficient. In this category of attacks we find most exploit attempts and network reconnaissance scans. In contrast, if an attack is expected to change how packets are produced and distributed in a network, \ac{ID2T} will only approximate the real effects of the attack. Most denial of service and botnet attacks fall into this category.

A second limitation exists that is related to the labeling of attack traffic. A dataset suitable for evaluating \ac{NIDS} needs to have labeled attacks. \ac{ID2T} indeed labels all synthetically injected attacks. However, if the provided background traffic contains attacks, not all attacks would be labeled. To try and leverage this problem, \ac{ID2T} analyses the background traffic provided by the user so as to highlight or make existing attacks stand out.

\subsection{Outline}
The remainder of this article is structured as follows.
In Section \ref{sec:requirements}, we propose a set of requirements for datasets that are suitable for the evaluation of \acp{NIDS} as well as for injecting synthetic attacks.
Section \ref{sec:related-work} discusses the state of the art in static and dynamic datasets for the evaluation of \ac{NIDS} and algorithms. We further categorize some of the defects found in these datasets.
With Section \ref{sec:toolkit} we begin the analysis of our toolkit (\ac{ID2T}), its architecture and components.
In addition, Section \ref{sec:tided} touches one important component of \ac{ID2T} that is responsible for the testing and analysis of intrusion detection datasets.
Subsequently, the attacks that can be generated by our toolkit are presented in Section \ref{sec:attacks}.
Section \ref{sec:evaluation}, shows use cases that demonstrate how \ac{ID2T} can be applied to evaluate anomaly and signature-based \acp{NIDS}.
Finally, Section \ref{sec:concl-future-work} concludes this article.


\section{Requirements}\label{sec:requirements}

Many requirements concern the creation and injection of synthetic attacks. On the one hand, there are requirements related to the tool that injects the attacks. On the other hand, there are requirements inherent to the resulting datasets, containing injected attacks, that are meant to be used for the evaluation of \acp{IDS}. This section emphasizes this distinction and divides the requirements into functional and non-functional ones.

\subsection{Requirements of Datasets Suitable for the Evaluation of IDSs}\label{sec:requirements-evaluation}

Irrespective of whether datasets contain synthetic or real data, they need to conform to certain requirements if they are to be used for the evaluation of \acp{NIDS}. From observations made in our research in the field of \ac{IDS} (i.e., \cite{vasilomanolakis2016phd,vasilomanolakis2016id2t,vasilomanolakis2015survey,Cordero2015id2t,vasilomanolakis2015skipmon}), as well as from the research of others \cite{kock2014datasets,shiravi2012toward,bhuyan2015,Mahoney2003a}, we have derived different requirements that make datasets usable in the context of evaluating \acp{NIDS}. In the following, these requirements are split into functional and non-functional ones.

\subsubsection{Functional Requirements}\label{sec:funct-requ}

These functional requirements focus on what is needed to construct or assemble datasets that enable different \acp{NIDS} to compare their performance against each other.

\begin{enumerate}
  \item Payload availability --- Due to privacy, network payloads are often unavailable. This limits the capability of \acp{NIDS} to detect attacks at the payload level (e,g., application layer exploits). To serve \acp{NIDS} that aim at detecting attacks at the payload level, one needs datasets that include payloads.
  \item Labeled attacks --- Datasets need labels that distinguish malicious from benign network traffic. Labels enable \acp{NIDS} to determine their detection accuracy and to establish a direct comparison against other \acp{NIDS}.
  \item Ground truth --- Labeled traffic is not enough if it cannot be guaranteed that the background traffic does not contain unlabeled attacks. Without ground truth, a direct comparison between different \acp{NIDS} is not possible.
  \item Growing --- A growing dataset is one that is constantly updated with traces of recent network traffic patterns and attacks.
  \item Attack diversity --- Datasets usually focus on certain types of attacks. A dataset suitable for the evaluation of different \acp{NIDS} requires a diversified set of network attacks, from \acp{DoS} to remote exploitation attempts.
\end{enumerate}

\subsubsection{Non-Functional Requirements}\label{sec:non-funct-requ}

The following non-functional requirements specify the criteria datasets need to satisfy to be of practical use.

\begin{enumerate}
  \item Public availability or ease of reproducibility --- Although being obvious, this requirement needs to be explicitly stated. Datasets need to be created with the goal of becoming public or easily reproduced. If a dataset is used for multiple evaluations, for the purpose of replicating experiments, it must be made public. If privacy is a concern, anonymizing the data should be carried out carefully so as to not introduce \emph{artifacts} (see Section~\ref{sec:artifacts} for a definition of the term). If the dataset cannot be public, reproducing it must be possible.
  \item Interoperability --- Network data needs to be shared using a common format. Although not a standard, the most commonly used format is the \ac{PCAP} file format.
  \item Good quality --- Through our empirical experience of generating synthetic attacks, as well as through the analysis of related work (see Section~\ref{sec:related-work}), we have identified diverse quality issues that need to be taken into account when synthetic attacks are injected into real traffic. These quality issues need to be addressed one by one, each one requiring special attention. In Section~\ref{sec:related-work}, we further discuss these issues (termed \emph{artifacts}) and categorize all observations of these in other datasets.
\end{enumerate}

\subsection{Requirements for Injecting Synthetic Attacks}\label{sec:requirements-inject}

Tools for injecting synthetic attacks need to be built with the goal of crating useful datasets for the evaluation of \acp{IDS}. These tools must be capable of working with arbitrary \ac{PCAP} files; the de facto standard for recording and sharing network communications. There should be no hard assumptions regarding the provenance of these files nor how they are generated. It is for this reason that considerable care needs to be taken when designing a tool that injects synthetic attacks into \ac{PCAP} files.

\subsubsection{Functional Requirements}\label{sec:funct-requ-1}

The following functional requirements have been derived from observations of how datasets are manually created. To alleviate the manual task of creating tailor-made labeled datasets with attacks, a tool that injects attacks needs to take several functional requirements into account. These next requirements focus on conferring tools with the ability to customize already existing attacks and create new ones without much additional effort.

\begin{enumerate}
  \item Packet level injection --- In order to evaluate any type of network \ac{IDS}, attacks must be injected at the lowest common denominator that all network \acp{IDS} can use, i.e., packets. Many \acp{IDS} directly use packets to perform intrusion detection. There is, however, another family of \acp{IDS} that work on higher abstraction levels, such as network flows~\cite{bhuyan2015}. If attacks are injected at the packet level, both types of \acp{IDS} can benefit from the generated datasets.
  \item Querying interface for network properties --- In order the generate realistic-looking synthetic attacks, an input \ac{PCAP} file needs to be analyzed to extract properties. These properties amount to quantitative characteristics such as \emph{average used bandwidth per host}, \emph{total number of packets sent by a specific host}, or \emph{the open ports observed for a given subnetwork}. These properties need to be made available to users that wish to inject attacks as well as to the program that creates the synthetic attacks. An \ac{API} is required to directly query the input files into which attacks will be injected. This enables, for example, a synthetic port scan to determine which ports, for a given host, were open in the input \ac{PCAP} file.
  \item Attack diversity --- Many network attack classifications exist. These classifications should be taken into account when providing tools for generating and injecting synthetic attacks. Rather than only focusing on a single type of attack, creating a diverse set of attacks is required. 
  \item Applicability to arbitrary \ac{PCAP} files --- As long as a \ac{PCAP} file contains network traffic alone, the file should be suitable for injecting synthetic attacks. To achieve this requirement, \ac{PCAP} files need to be analyzed to replicate their properties onto injected attacks. For instance, if a network only contains local traffic (from the 192.168.1.0 subnet), by default attacks should not originate from IP addresses outside this subnet (unless this is desired and manually configured).
  \item Headless operation --- Working with arbitrary \ac{PCAP} files implies that their size is unbounded. Packet captures of crowded networks can easily reach a size of hundreds of gigabytes. Analyzing and injecting attacks into these files might take considerable time and memory resources. It is assumed, therefore, that cluster environments or headless systems are used to process large files. For this reason, tools that inject attacks need to work in headless environments with small or no user interaction.
  \item Modeling of packet behavior and payload --- Attack injection tools need to model attacks at the packet and payload levels. Attacks at the packet level are more concerned about the quantity of the sent packets rather than on the payload. Port scans and \ac{DDoS} attacks are examples of these. In contrast, attacks at the payload level may carefully craft the data sent by each packet. Exploit attacks against a web server are examples of these. Some attacks are concerned with both levels, such as when creating synthetic botnet generated traffic. This traffic sometimes contains important payload information that defines the attack, other times, the traffic is used as a \ac{DDoS} attack in which the payload is not important.
\end{enumerate}

\subsubsection{Non-Functional Requirements}\label{sec:non-funct-requ-1}

The following non-functional requirements are needed to enable synthetic attack injection tools to process arbitrary \ac{PCAP} files. The main challenge when not being limited to certain \ac{PCAP} files is to consider the different scenarios from where these files may originate. Home and office environments generate limited traffic belonging to only few sub-networks. In contrast, large enterprise or university networks may generate massive quantities of data with heterogeneous characteristics.

\begin{enumerate}
  \item Scalability --- When mimicking the properties of network data onto a synthetic attack, many different statistics need to be collected. Calculating statistics on top of network data is typically demanding both in terms of memory and processing resources. Tools that inject synthetic attacks need to take this into account and calculate statistics as efficiently as possible. This guarantees the scalability of the tool when large \ac{PCAP} files are analyzed and injected with synthetic attacks.
  \item Extensibility --- New attacks are developed each day. A set of tools and libraries are needed to easily and rapidly model these new attacks for injection.
  \item Usability --- One of the main reasons static datasets are widely used is due to their ease of use. Static datasets have many disadvantages: they never change and become easily outdated. Their advantages, however, outweigh the disadvantages: they can be used immediately without extensive setup. Tools that generate either real or synthetic traffic often require complicated hardware and setups. To be as useful as static datasets, tools that inject synthetic attacks need to be usable. The user should be involved as little as possible in the injection process but have a detailed control over the injections.
  \item Open source and public availability --- Researchers often build custom tools to evaluate their work without making the tools available in open platforms. It is also common that tools are lost when researchers change subject areas or institutions. For example, the tools proposed by Shiravi et al.~\cite{shiravi2012toward} and Brauckhoff et al.~\cite{Brauckhoff2008} can no longer be easily found online although they used to be available.
\end{enumerate}


\section{Related Work and Defect Analysis}
\label{sec:related-work}

\begin{figure}[h!]
	\center{\includegraphics[width=0.9\textwidth]
		{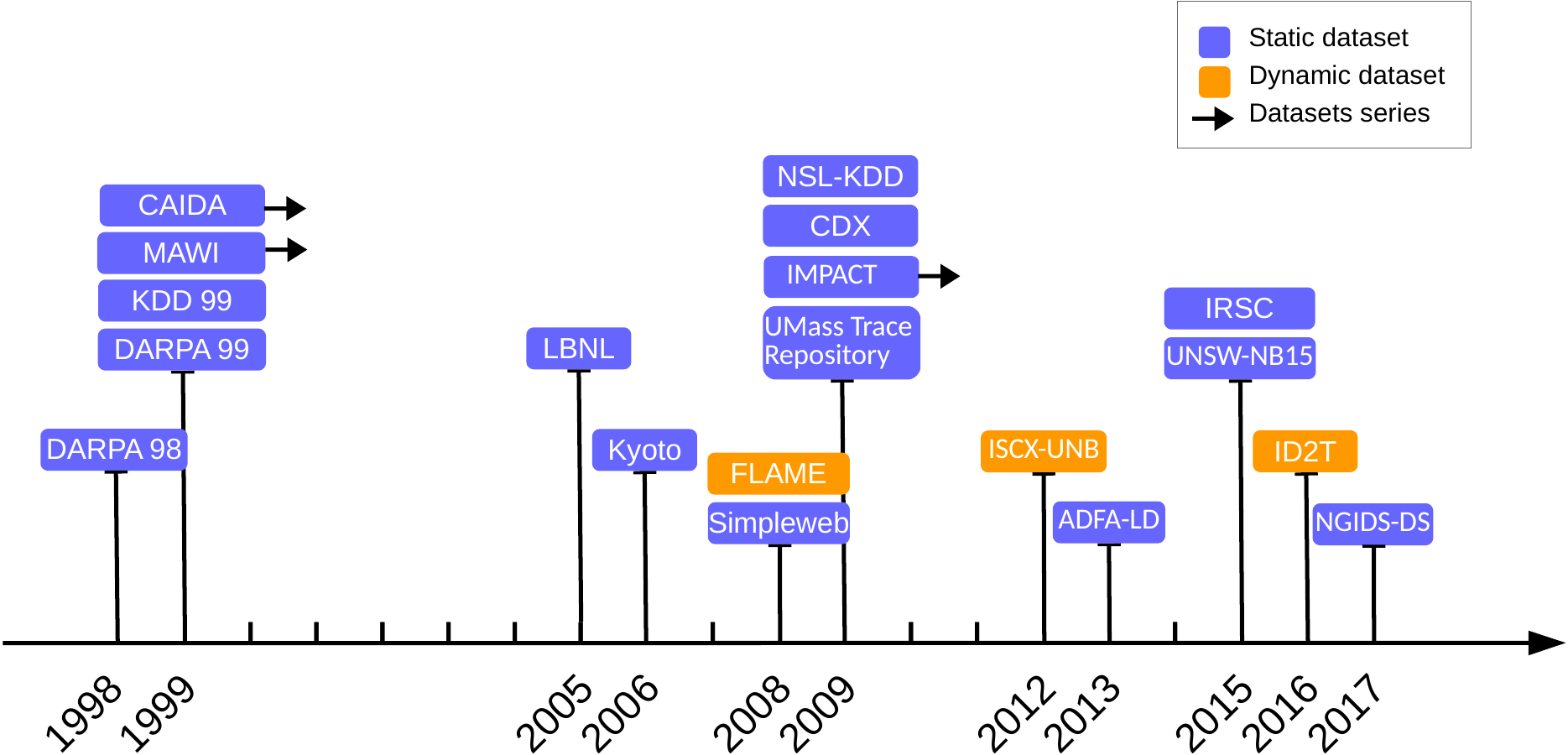}}
	\caption{\label{fig:dss_timeline} A historical timeline overview of datasets intended for intrusion detection.}
\end{figure}

A good quality dataset is one that allows researchers to identify the ability of an \ac{IDS} to detect anomalies \cite{bhuyan2015}, preferably allowing to draw valid conclusions about the appropriateness of the \ac{IDS} (its efficiency, accuracy, validity scope, etc). Although many of the available datasets nowadays are valuable to the research community, they unfortunately fail to fulfill all the requirements proposed in Section \ref{sec:requirements}. The existing datasets can be classified into two categories: \textit{static} datasets and \textit{dynamically generated} datasets (see Figure \ref{fig:dss_timeline} for a historical overview). In this section, first, we provide basic information about some of the most popular static datasets. Afterwards, we present a description about the dynamically generated datasets. Lastly, this section discusses the deficiencies and artifacts that can be found in datasets.

\subsection{Static Datasets}
A static dataset is a dataset that was generated once during a limited period of time. Such a dataset can be collected from a real-world network or it can be generated synthetically. Currently, there are several static datasets available for \ac{IDS} evaluation purposes. Unfortunately, research suggests that these datasets are not adequate for mainly two reasons. First, many of them are out of date. They were created many years ago based on old versions of protocols, old applications, and attacks that exploit old vulnerabilities. Thus, these datasets are not realistic datasets at the present time. Second, these datasets contain many defects that indicate synthetic generation. In the following, we briefly discuss some well-known static datasets. It is worth to mention that similar datasets can be found in other fields, such as the \ac{ICS} field, where building testbeds is a challenging task \cite{sugumar2017testing, holm2015survey}. However, the testbeds in such fields are built under application-specific conditions, and are therefore considered out of the scope of this article.

\subsubsection{DARPA $99$}
The DARPA $99$ is one of the most well-known datasets in the field. In fact, DARPA $99$ is an upgraded version of the DARPA $98$ dataset. 
The two datasets were collected during two intrusion detection evaluations \cite{cunningham1999evaluating, Lippmann2000}.
A testbed was developed to generate normal and malicious traffic. The normal traffic tended to be similar to that seen between a military base and the Internet. A custom software automaton was used to simulate hundreds of users running UNIX applications. 
In DARPA $98$, the included attacks cover four categories, namely, \ac{DoS}, \ac{R2L}, \ac{U2R}, and Surveillance/Probing. The traffic was collected for seven weeks with labeled attacks and two weeks with unlabeled attacks. 
In DARPA $99$, Windows NT hosts were included in the evaluation as victims and more attacks types were covered. 
The dataset consists of two weeks of normal traffic with no attacks, one week with a few labeled attacks, and two weeks of unlabeled traffic. 
DARPA $98-99$ datasets were highly innovative for their time and had been widely used for \ac{IDS} evaluation. However, later on, several studies have marked these datasets as unreliable due to a number of limitations and issues (see Section \ref{sec:artifacts}). 

\subsubsection{KDD $99$}
This dataset was created on the basis of the DARPA $98$ during a competition for network intrusion detectors \cite{kdd99task}. KDD $99$ contains data records, rather than packets, where each record contains $31$ features that describe a connection. A connection is a sequence of TCP packets that flow between two defined IP addresses, under defined protocol and within defined time interval. The raw data of DARPA $98$ was used to extract seven million connection records. All connections were labeled as either normal or malicious with one specific attack type.
This dataset was used heavily to evaluate \acp{IDS} at that time. Later on, several shortcomings in this dataset were pointed out by various studies. 

\subsubsection{MAWI}\label{sec:mawi_dataset}
The \ac{MAWI} dataset corresponds to a family of datasets that are collected from an operational testbed network, connecting Japanese universities and research institutes with various networks all over the world \cite{sony2000traffic}. The traffic is stored in the \textit{tcpdump} raw format, so the header information is available and can be used for further analysis. However, for privacy reasons, IP addresses are scrambled and packets' payloads are removed. Unfortunately, MAWI datasets are not labeled, thus, there is no ground truth to be used in \ac{IDS} evaluation.
Fontugne et al. \cite{Fontugne2010} proposed a method to label the anomalies in a subset of MAWI datasets. The labeled dataset is known as MAWILab dataset. The labels are obtained using an advanced graph-based methodology that compares and combines several independent anomaly detectors. The dataset is updated constantly to include new traffic and anomalies from upcoming applications.
Nevertheless, as the detectors are anomaly-based, false positives are expected to be present in the dataset.

\subsubsection{NSL-KDD}
Tavallaee et al. \cite{Tavallaee2009} conducted a statistical analysis on the KDD $99$ dataset and found two deficiencies. First, more than $75\%$ of the data records are duplicated, which makes the learning algorithms less sensitive for infrequent records, which can be malicious. 
Second deficiency is the low level of difficulty. Using typical classifiers, the authors were able to label the dataset correctly with an unreasonably high accuracy rate. 
In NSL-KDD, the duplicated records were removed and the difficulty level was increased by selecting records out of KDD $99$ based on the inverse of classification accuracy rate. 

\subsubsection{CDX}
This dataset was collected from the \ac{CDX} $2009$ competition \cite{Sangster2009}. 
The knowledge about the locations of the defenders and the attackers in the network was used to label the traffic as normal or malicious.
In contrast to the DARPA $98-99$, the CDX dataset is more realistic, since it presents the live traffic of real human activities. Nevertheless, the ratio between the normal and malicious traffic was almost equal, not representing real world observations. In addition, the dataset has a small volume because of the limited duration of the competition. This is problematic for anomaly detectors that require a long training period. 

\subsubsection{Other Datasets}
In the following, we briefly discuss other \ac{IDS} public datasets. An overview of static datasets and their properties can be found in Table \ref{tab:comparison} \cite{zuech2015new, Sangster2009, barbosa2010simpleweb, kock2014datasets, shiravi2012toward}. 

\begin{itemize}
	\item \textbf{CAIDA}: \ac{CAIDA} collects data at different locations, and provides this data to the research community, taking into account the privacy of individuals and organizations who participate in generating the data \cite{caida}.
	
	\item \textbf{LBNL}: \ac{LBNL} collected packet traces for more than $100$ hours of internal enterprise traffic, and released this data publicly in an anonymized form \cite{lbnl2005}.
	
	\item \textbf{Simpleweb}: A dataset was collected, Twente university network, via a honeypot in September $2008$. Several network services were hosted on that honeypot, which was directly connected to the Internet. The honeypot only captured suspicious traffic \cite{barbosa2010simpleweb}.
	
	\item \textbf{IMPACT}: The \ac{IMPACT} project provides an open platform for dataset exchange. This platform connects producers and consumers of cyber-risk-relevant data. The existing data was collected by various organizations and universities \cite{impact}.
	
	\item \textbf{UMass Trace Repository}: UMass university provides a trace repository that contains network, storage, and other data traces. The data was extracted during various experiments and hence is specialized; it reflects specific network behavior and attacks that were captured for experimental purposes \cite{umass2009}. 
	
	\item \textbf{ADFA-LD}: This dataset is provided by the university of New South Wale and consists of normal and malicious Linux based system call traces. Only host logs had been collected \cite{haider2017generating}.
	
	\item \textbf{IRSC}: The dataset, created by the \ac{IRSC}, consists of packet captures and network flow data and includes labels \cite{zuech2015new}. 	
	
	\item \textbf{UNSW-NB15}: This dataset contains real modern normal traffic and synthetic attacks. It was specifically generated over a commercial penetration testing environment \cite{Moustafa2015,haider2017generating}.
	
\end{itemize}

\begin{table}[h]\centering
	\begin{threeparttable}
		\begin{tabular}{lcccccc}
			\toprule
			\thead{Dataset} & \thead{Availability} &  
			\thead{Synthetic} & \thead{Payload} &
			\thead{Ground\\truth} &  
			\thead{Labeled\\attacks} & \thead{Updates} \\
			\midrule	
			DARPA 98-99 & \textcolor{olive}{\checkmark} & \textcolor{olive}{\checkmark} & \textcolor{olive}{\checkmark} & \textcolor{olive}{\checkmark} & \textcolor{olive}{\checkmark} & \textcolor{red}{\sffamily x} \\
			KDD 99 & \textcolor{olive}{\checkmark} & \textcolor{olive}{\checkmark} & \textcolor{olive}{\checkmark} & \textcolor{olive}{\checkmark} &  \textcolor{olive}{\checkmark} & \textcolor{red}{\sffamily x} \\
			MAWILab & \textcolor{olive}{\checkmark} & \textcolor{red}{\sffamily x} & \textcolor{red}{\sffamily x} & \textcolor{red}{\sffamily x} & \textcolor{olive}{\checkmark} & \textcolor{olive}{\checkmark} \\
			CAIDA & \textcolor{olive}{\checkmark}$^1$ & \textcolor{red}{\sffamily x} & \textcolor{olive}{\checkmark} & \textcolor{red}{\sffamily x} & \textcolor{red}{\sffamily x} & \textcolor{olive}{\checkmark} \\
			SimpleWeb & \textcolor{olive}{\checkmark} & \textcolor{red}{\sffamily x} & \textcolor{red}{\sffamily x} & \textcolor{red}{\sffamily x} & \textcolor{red}{\sffamily x} & \textcolor{red}{\sffamily x} \\
			NSL-KDD &  \textcolor{olive}{\checkmark} & \textcolor{olive}{\checkmark} & \textcolor{olive}{\checkmark} & \textcolor{olive}{\checkmark} & \textcolor{olive}{\checkmark} & \textcolor{red}{\sffamily x} \\
			CDX & \textcolor{red}{\sffamily x} & \textcolor{olive}{\checkmark} & \textcolor{olive}{\checkmark} & \textcolor{olive}{\checkmark} &  \textcolor{olive}{\checkmark} & \textcolor{red}{\sffamily x} \\
			IMPACT & \textcolor{olive}{\checkmark}$^2$ & \textcolor{olive}{\checkmark} & \textcolor{red}{\sffamily x} & \textcolor{red}{\sffamily x} & \textcolor{red}{\sffamily x} & \textcolor{olive}{\checkmark} \\
			UMass & \textcolor{olive}{\checkmark} & \textcolor{olive}{\checkmark}$^3$ & \textcolor{olive}{\checkmark} & \textcolor{red}{\sffamily x} & \textcolor{red}{\sffamily x} & \textcolor{red}{\sffamily x} \\
			RIPE & \textcolor{olive}{\checkmark}$^4$ & \textcolor{red}{\sffamily x} & \textcolor{olive}{\checkmark}$^5$ & \textcolor{red}{\sffamily x} & \textcolor{red}{\sffamily x} & \textcolor{olive}{\checkmark}$^6$ \\
			IRSC & \textcolor{red}{\sffamily x} & \textcolor{olive}{\checkmark} & \textcolor{olive}{\checkmark} & \textcolor{olive}{\checkmark} & \textcolor{olive}{\checkmark} & \textcolor{red}{\sffamily x} \\
			UNSW-NB15 & \textcolor{olive}{\checkmark} & \textcolor{olive}{\checkmark}$^7$ & \textcolor{olive}{\checkmark} & \textcolor{olive}{\checkmark} & \textcolor{olive}{\checkmark} & \textcolor{red}{\sffamily x} \\
			\bottomrule	
		\end{tabular}
		\begin{tablenotes}
			\scriptsize
			\item 1 Access to some datasets requires special permissions. \hspace{0.2cm} 5 Not available in all datasets.
			\item 2 Access is limited to U.S. researchers.\tab\tab6 Not updated regularly.
			\item 3 A variety of datasets, most of them synthetic.  \tab 7 Real benign traffic and synthetic attacks.
			\item 4 Registration is required.
		\end{tablenotes}
		\caption{\label{tab:comparison} Comparison of static datasets.}
	\end{threeparttable}
\end{table}


\subsection{Dynamically Generated Datasets}
Network behavior and traffic patterns change continuously and attacks evolve rapidly. This created a need for datasets that are modifiable, extensible, and reproducible. Therefore, researchers have proposed tools that are capable of generating synthetic datasets dynamically. The main idea of these tools is to consider the contemporary traffic characteristics and replicate them in synthetic traffic, which in turn emphasizes the realism of the generated datasets. In this section, we present the two existing dynamic datasets, followed by a brief comparison of them in Table \ref{tab:dynmaic_DSs}.

\subsubsection{FLAME}
Bauckhoff et al. \cite{Brauckhoff2008} proposed FLAME, an application that generates and injects parameterized anomalies into a given flow trace.
Three classes of anomalies are offered by FLAME: (\textit{i}) Additive anomalies where synthetic flows are added to a background trace, but without interacting with the existing flows, e.g., network scans; (\textit{ii}) Subtractive anomalies in which selected flows are removed  from background traffic, e.g., ingress shifts; (\textit{iii}) Interactive anomalies where synthetic flows are added to a background trace, and  selected flows are removed, e.g., \ac{DoS} attacks that cause a network congestion. In particular, three use cases were implemented: a TCP SYN network scan, ingress shifting, and TCP SYN \ac{DoS} attack. 


The usage of network flows, rather than payloads, as background traffic makes datasets generated by FLAME usable only for flow-based intrusion detection algorithms. In addition, many attacks cannot be detected by only using network flows. In this context, FLAME is capable of creating a limited range of attacks which have a flow footprint. Moreover, the implementation of FLAME is not publicly available anymore due to its discontinuation. 

\subsubsection{ISCX-UNB}
Shiravi et al. \cite{shiravi2012toward} proposed a mechanism to generate dynamic datasets based on the concept of profiles (i.e., abstract representations of certain features and events in network traffic). Two classes of profiles were proposed, namely, $\alpha$ and $\beta$. The $\alpha$ profiles describe attack behavior and are used to generate malicious traffic. The $\beta$ profiles represent legitimate agents' activities and are used to generate normal traffic.

Real network traffic was analyzed to extract mathematical distributions of various properties, such as packet sizes, sizes of payloads, etc. These distributions were encapsulated in $\beta$ profiles to create models of normal traffic. Several $\beta$ profiles were created to simulate different legitimate agents that use various network protocols.
The $\alpha$ profiles were created manually based on human knowledge. Each profile represents an attack scenario and a number of attacks were implemented. 
A testbed network was built and a combination of $\alpha$ and $\beta$ profiles were executed to generate traffic. 

The quality, of the profile models in this work, however, raises questions. The $\beta$ profiles were created based on real traffic that was not verified to be free of attacks, and this degrades the confidence degree in the correctness of the profiles. Moreover, human assistance is required to execute $\alpha$ profiles, which leads to variances on how attacks are carried out. Thus, the reproducibility of the generated dataset is reduced. 
Lastly, the authors provide only the generated dataset to the research community, but not the implemented profiles.

\begin{table}[h]\centering
	\begin{threeparttable}
		\begin{tabular}{lcccccc}
			\toprule
			\thead{Dataset} & \thead{Availability} &  
			\thead{Synthetic} & \thead{Payload} &
			\thead{Ground\\truth} &  
			\thead{Labeled\\attacks} & \thead{Tool\\updates} \\
			\midrule	
			FLAME & 
			\textcolor{red}{\sffamily x}  & \textcolor{olive}{\checkmark} & 
			\textcolor{red}{\sffamily x} &
			\textcolor{red}{\sffamily x}$^1$ & 
			\textcolor{olive}{\checkmark} & 
			\textcolor{red}{\sffamily x} \\
			ISCX-UNB & 
			\textcolor{red}{\sffamily x}$^2$ & \textcolor{olive}{\checkmark} & \textcolor{olive}{\checkmark} & \textcolor{olive}{\checkmark} &  \textcolor{olive}{\checkmark} & 
			\textcolor{red}{\sffamily x} \\
			\bottomrule	
		\end{tabular}
		\begin{tablenotes}
			\scriptsize
			\item 1 Users are responsible for providing benign traffic as an input.
			\item 2 One output dataset is available but not the tool. 
		\end{tablenotes}
		\caption{Comparison of dynamic datasets.}
		\label{tab:dynmaic_DSs}
	\end{threeparttable}
\end{table}


\subsection{Defect Classification}\label{sec:artifacts}

In this section, we expose and categorize defects, that we as well as other researchers have detected. \ac{ID2T} actively tries to avoid committing the same mistakes discovered in other datasets, either static or dynamic. We use the term \emph{deficiency} to refer to a problem that appears in a static dataset. In contrast, we use the term \emph{artifact} to indicate a problem that has appeared as a side-effect of creating synthetic network traffic. \emph{Artifacts} are associated to dynamic datasets only.
Synthetic data is any data that is not acquired through direct measurements~\cite{McGraw-Hill_Dict2009}. In the context of networks, synthetic data refers to network traffic that is not captured directly from real-world networks, instead, it is generated by software to mimic real-world traffic.

\subsubsection{Categorization of Dataset Defects}\label{sec:artif-class}

\begin{figure}[ht]
	\centering
	\includegraphics[width=0.8\textwidth]{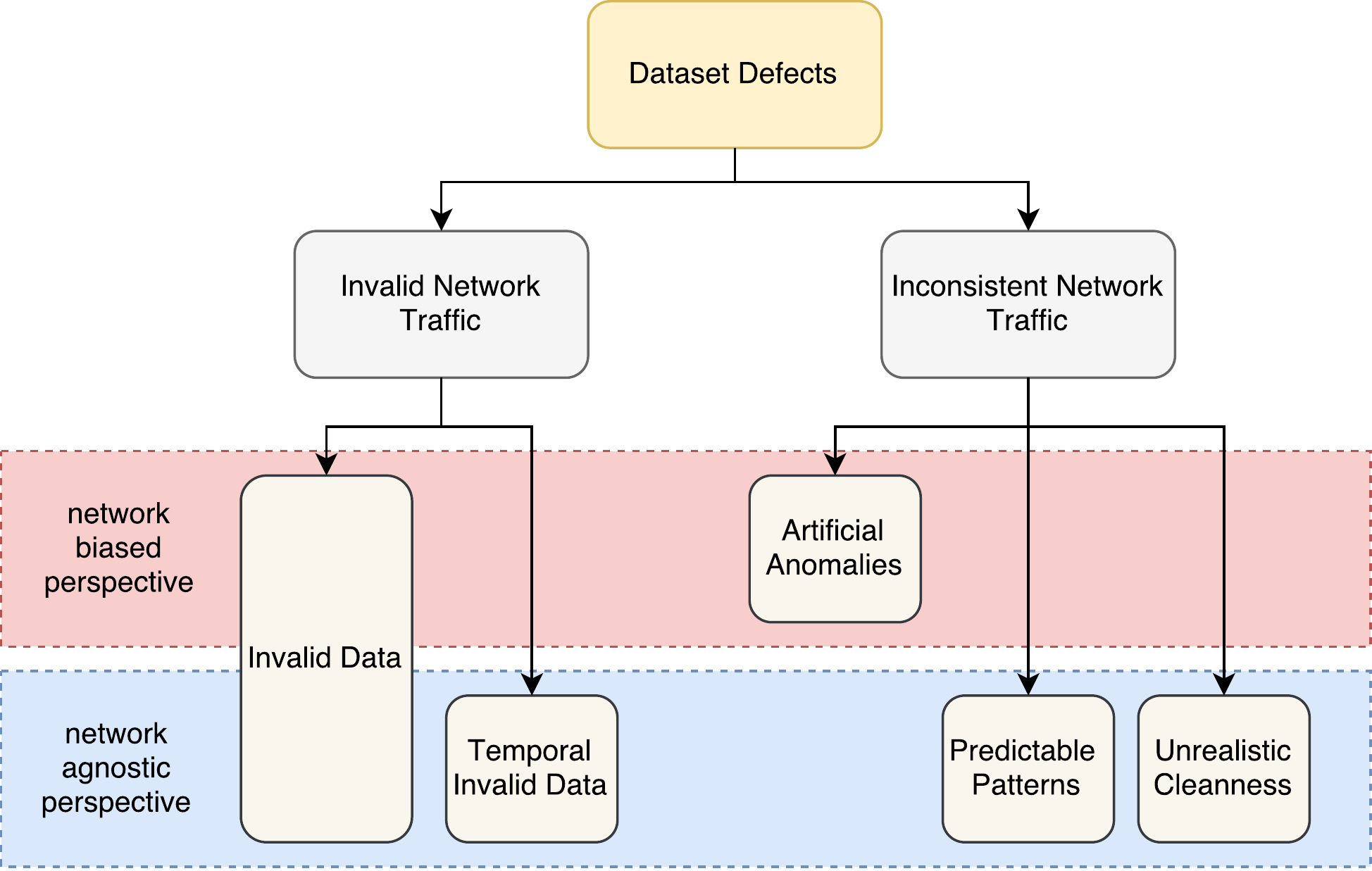}
	\caption{Categories of defects found in static or dynamic datasets.}%
	\label{fig:defects_categories}
\end{figure}

We have categorized the dataset defects according to Figure~\ref{fig:defects_categories}. Dataset defects can be divided in two classes, namely \emph{Invalid Network Traffic} and \emph{Inconsistent Network Traffic}. Invalid Network Traffic defects are problems that relate to the incorrect usage of network protocols or specifications. Inconsistent Network Traffic defects are those pertaining the creation of inadvertent traffic patterns or network packets. These two defect classes are characterized by different symptoms depending on two different perspectives. From a \emph{Network Biased Perspective}, certain traffic presents defects only when the background traffic is taken into account; otherwise, it looks normal. From a \emph{Network Agnostic Perspective}, traffic patterns present defects regardless of the background traffic. In the following, each piece of the classification is explained and the defects found are classified.

\begin{enumerate}
  \item \textbf{Invalid Network Traffic}:  Traffic patterns or individual packets that are invalid because they contradict a network protocol specification. We encountered two reasons concerning these class of defects, namely \emph{Invalid Data} and \emph{Temporary Invalid Data}. 
        \begin{enumerate}
          \item \textbf{Invalid Data}: Invalid traffic according to a protocol specification. These defects have a slight change of meaning depending on the perspective. From a network specific perspective, traffic data in a dataset is invalid if the characteristics and physical limitations of a network (captured by the dataset) are not respected. For example, if packets have a larger \ac{MTU} than what the network hardware supports; also, if packets use more bandwidth than what is available. These two examples are cases of \emph{artifacts}. From a network agnostic perspective, traffic data is invalid if protocols are violated that should or cannot be. For example, any TCP packet using port zero.
          \item \textbf{Temporary Invalid Data}: Traffic data that used to be valid according to an old protocol but are not anymore. These defects are only relevant when looked from a network agnostic perspective. An example of such a defect is found in the DARPA $99$ dataset \cite{Lippmann2000}. Here, the IPv4 \ac{ToS} field uses an old standard found in RFC $791$~\cite{postel1981rfc}. The \ac{ToS} field has a different meaning according to the more recent RFC $3260$~\cite{rfc3260}.
        \end{enumerate}				
  \item \textbf{Inconsistent Network Traffic}: Network traffic that does not reflect the same characteristics of some background network traffic. Inconsistencies can originate from wrong assumptions of how a network behaves or unforeseen  problems in the process of generating synthetic traffic.
        \begin{enumerate}
          \item \textbf{Artificial Anomalies}: Anomalous data patterns inadvertently added to network traces. These can result from either the generation of synthetic attacks or the incorrect set-up of hardware. As an anomalous pattern depends on the specific network, this is a defect found only from a network biased perspective. For example, Mahoney et al.~\cite{Mahoney2003} were able to detect $33$ attacks in the DARPA $99$ dataset \cite{Lippmann2000} due to the use of the same \ac{TTL} values in the packets of all attacks.
          \item \textbf{Predictable Patterns}: Patterns that appear repeatedly such that network traffic diversity is constricted, making the traffic incorrectly correlate with bogus information. These defects are not expected in any network and are therefore taken into account from a network agnostic perspective. As an example, Brauckhoff et al.~\cite{Brauckhoff2008} generated synthetic SYN port scans with this defect. The authors, without knowing the repercussion of doing so, used the same delay between the reply packets of the attack. This led to a predictable pattern of inter-arrival times that highly correlate with the attack. Similarly, the DARPA $99$ dataset~\cite{Lippmann2000} utilized \ac{TTL} values with only nine different values across the entire dataset, making some values highly correlate with the attacks alone.
          \item \textbf{Unrealistic Cleanness}: An excessively clean dataset can be a sign of network traffic that does not have the characteristics of real-world network traffic. Most networks will periodically observe issues such as incorrect packet checksums or TCP packet retransmissions and duplication. From a network agnostic perspective, a complete lack of these issues signals a dataset that does not conform to a real-world and live network. 
        \end{enumerate}
\end{enumerate}
We now go on to describe our dataset generation framework followed by the tools and applications of it.


\section{The Intrusion Detection Dataset Toolkit (ID2T)}
\label{sec:toolkit}
The \acf{ID2T} is software that generates and injects synthetic attacks into a \ac{PCAP} file. The main aim of \ac{ID2T} is to dynamically create high quality datasets for \ac{IDS} evaluation purposes. \ac{ID2T} was designed to provide the research community with datasets that meet the aforementioned requirements of \ac{IDS} datasets (see Section \ref{sec:requirements-evaluation}). At a glance, \ac{ID2T} takes a \ac{PCAP} file as input and provides a new \ac{PCAP} file as output; the latter contains the original input traffic along with synthetic attacks.

The main architecture of \ac{ID2T} was presented in \cite{vasilomanolakis2016id2t,Cordero2015id2t}. In this work, we extend the \textit{Attacks} module to include a wider range of attacks. Generating additional attacks with different characteristics requires considering further statistical properties of the background traffic, therefore, we extend as well the \textit{Statistics} module to cover this aspect. In addition, we develop a new module \textit{\ac{TIDED}}, which analyzes the input traffic properties in order to indicate potential problems.
In this section, we provide a detailed description of \ac{ID2T} and its extended architecture.

\subsection{ID2T Architecture} \label{sec:toolkit:architecture}
\label{sec:id2t-architecture}

\begin{figure}[h]
  \centering
  \includegraphics[width=\textwidth]{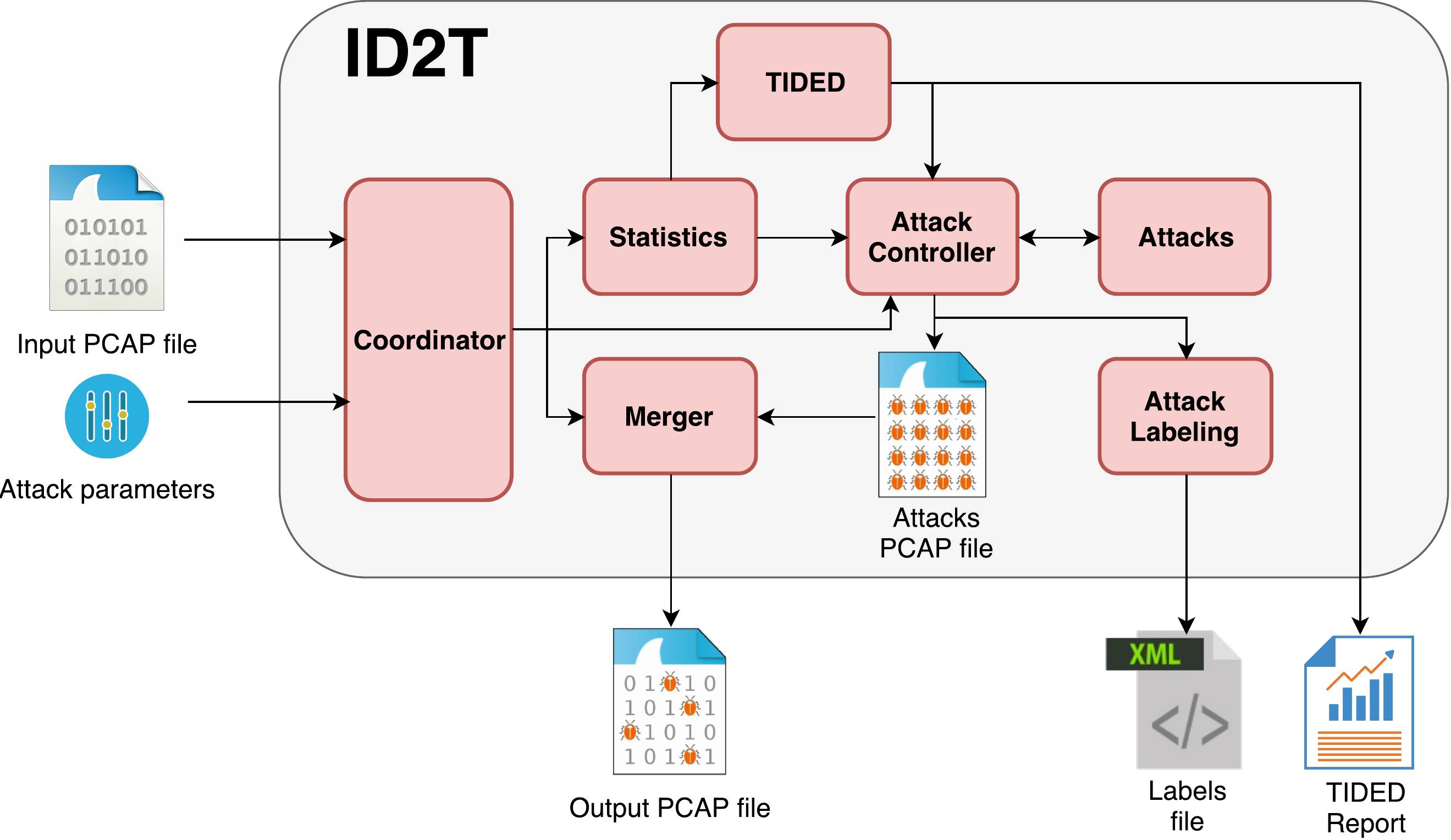}
  \caption{An overview of the ID2T architecture}
  \label{fig:architecture}
\end{figure}

Figure \ref{fig:architecture}, illustrates the architecture of \ac{ID2T} in which arrows represent flow of data between the computation modules. In the following, we describe in detail the architecture by first focusing on the input and the output of \ac{ID2T} and subsequently on the core components of the system.

\subsubsection{Input}
\ac{ID2T} receives two inputs from the user: a \ac{PCAP} file and the attack parameters.
As mentioned in Section \ref{sec:introduction}, the main assumption behind \ac{ID2T} is that the user is responsible for the input \ac{PCAP} file. That is, the \ac{PCAP} may or may not contain attack data.
In both cases, the toolkit replicates the input traffic characteristics in the synthetic attacks.
Hence, \ac{ID2T} does not examine the input for signs of attack data. It does, however, provide a test module that conducts a quantitative analysis of the input \ac{PCAP} file. This analysis produces several
statistical properties, which can be used to infer potential problems in the input traffic (e.g., incorrect TCP checksums or unexpected values in packet headers).  
The second input is the attack parameters, which specify the properties of the synthetic attack. Examples of such parameters can be the IP addresses of the attacker and the victim, the utilized ports, and the location in which the attack will be injected in the input traffic.

\subsubsection{Output}
As depicted in Figure \ref{fig:architecture}, the output of the system is threefold. 
First, \ac{ID2T} creates a \ac{PCAP} file that includes the input traffic along with the synthetic attacks. 
Second, the toolkit can (optionally) perform a multitude of statistical tests to explore the characteristics of the input traffic. The results of the tests are presented in a set of graphs and a text file denoted "TIDED report" in Figure \ref{fig:architecture}.
Finally, \ac{ID2T} generates a labels file that contains information about the injected attacks and their locations in the output \ac{PCAP} file.

\subsubsection{ID2T-Modules}

In the following, we present the core modules of the system, namely the \textit{Coordinator}, the \textit{Statistics}, the \textit{Attack Controller}, the \textit{Merger}, the \textit{Attack Labeling} and \textit{\ac{TIDED}}. 

\paragraph{Coordinator} This module initiates the process, triggers other modules, and passes them the required parameters. More specifically, the \textit{Coordinator} receives the path of the input \ac{PCAP} file and the attack parameters from the user, and forwards these two inputs to the \textit{Statistics} and the \textit{Attack Controller} modules, respectively. 

\paragraph{Statistics} This module collects statistical properties of the input \ac{PCAP} file. The module is also responsible for calculating the hash \footnote{SHA-224 hash function} of the input \ac{PCAP}, which is used to check whether the file was previously loaded and analyzed. In this case, the recalculation of the file's properties is not required. Otherwise, the \textit{Statistics} module analyzes the packets of the file sequentially and collects four types of features. 

First, properties of the file as a whole are collected and taken into account; these can be features such as the total packet count, the capture duration and the average packet size. Second, properties that relate to each host in the network, such as the sent and received packets are determined. Third, the distribution of some header fields, e.g., the protocols, the ports , the \acp{TTL}, and the window sizes are calculated. Finally, features with regard to the TCP connections, such as the average packet rate and the packet inter-arrival time are taken into account.
The aforementioned properties are mainly used by the \textit{Attack Controller} module to guarantee that the synthetic attacks exhibit similar characteristics to the input traffic when this is expected from an attack that would be created in real world. In addition, the statistical properties provide the main input for the traffic analysis that is conducted by the \textit{\ac{TIDED}} module.

\paragraph{Attack Controller} The main tasks of this module is to validate the attack parameters provided by the user, and to call the corresponding \textit{Attacks} module to generate the synthetic attack. The \textit{Attack Controller} module provides the \textit{Attacks} module with the statistical properties of the input \ac{PCAP} file. Afterwards, it produces a temporary \ac{PCAP} file that contains the generated attack. 

\paragraph{Attacks} This module contains the attacks that \ac{ID2T} can generate and inject. This module was designed to be extensible, in the sense that a user can implement and add their own attacks. Currently, the module contains four classes of attacks, namely, probing and surveillance, exploits, resource exhaustion, and botnet infection. For each attack, this module receives and processes a different set of parameters and, in some case, additional input files. An in-depth discussion of the attacks that are supported by \ac{ID2T} is given in Section \ref{sec:attacks}.

\paragraph{Merger} After generating a temporary attack \ac{PCAP} file by the \textit{Attack Controller} module, the \textit{Merger} module injects the attacks by merging that file with the input \ac{PCAP} file according to the timestamps of the packets.

\paragraph{Attack Labeling} The task of this module is to create a file that contains labels of the injected attacks. The \textit{Attack Controller} module sends information about the injected attacks to the \textit{Attack Labeling} module, which in turn writes it in XML format into a text file. The information includes the names of the attacks and the timestamps where they start and end in the output \ac{PCAP} file.	

\paragraph{TIDED} This module provides users with a quantitative analysis of the input traffic characteristics. A detailed description of the \textit{\ac{TIDED}} module is given in Section \ref{sec:tided}.

\subsection{Modules' Performance} 
This section presents an overview of the performance of two modules, namely, the \textit{statistics} and the \textit{attacks} modules. Our experiments show that the ability of these modules to generate the internal data needed for \ac{ID2T} is acceptable from a timing perspective, meaning that no bottlenecks have so far been discovered in running the modules when feeding each other with data. In the following, we show two charts: the statistics generation performance indicator, and the \ac{DDoS} attack data creation indicator.

\begin{figure}[ht]
	\center{\includegraphics[clip, trim= 0cm 3cm 0cm 3.5cm, width=0.8\textwidth] {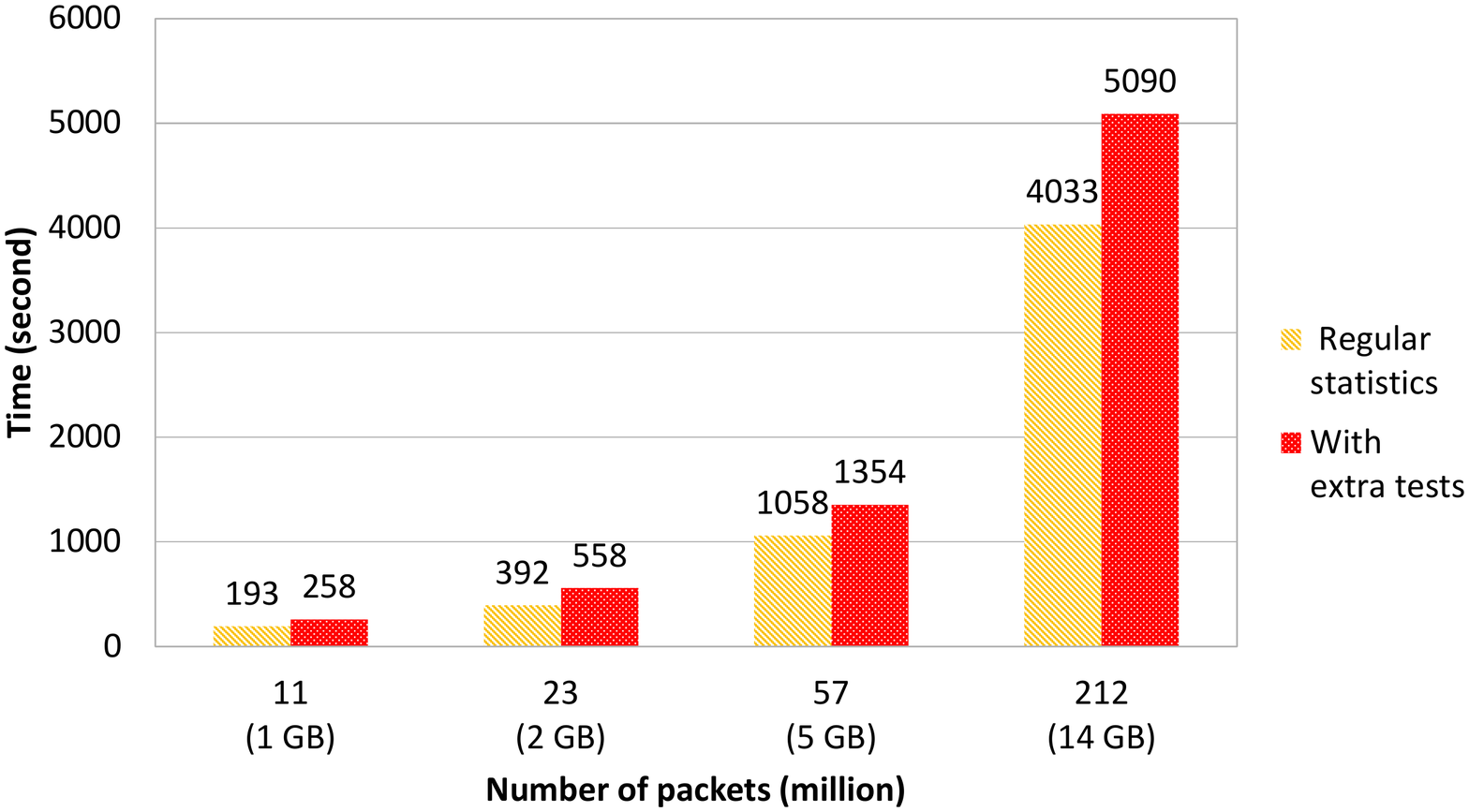}}
	\caption{Performance of collecting statistics}
	\label{fig:id2t-statistics-collecting}
\end{figure}

\paragraph{a) Generation of Statistics:} One of our proposed requirements (see Section \ref{sec:requirements}) is the ability to handle arbitrary files. Hence, it is important to examine how ID2T's architecture handles large files. The only bottleneck to take into account, for handling large files, is located in the \textit{statistics} module. Figure~\ref{fig:id2t-statistics-collecting} shows the time required to handle files of different sizes. The \ac{PCAP} files utilized in the experiment are from the MAWI archives \footnote{http://mawi.wide.ad.jp/mawi/samplepoint-F/2015/201506021400.html}. In particular, four different files are examined, ranging from approximately $11$ million to $212$ million packets. For each file, two different measurements were examined. First, for collecting the so-called regular statistics, which encompass the features that were presented in the \textit{Statistics} module description. Second, measurements are taken for further data analysis. These extra tests include, but are not limited to, calculating IP addresses' entropies, examining the correctness of TCP checksums, and checking the availability of packet payloads. These tests are parts of the \textit{\ac{TIDED}} module, and are further explained in Section~\ref{sec:tided}. At a glance, Figure~\ref{fig:id2t-statistics-collecting} suggests that moderate network files, e.g., a 14GB PCAP file, can be analyzed by \ac{ID2T} in a reasonable time frame.

\begin{figure}[ht]
	\center{\includegraphics[clip, trim= 0cm 3cm 0cm 3.5cm, width=0.7\textwidth]
		{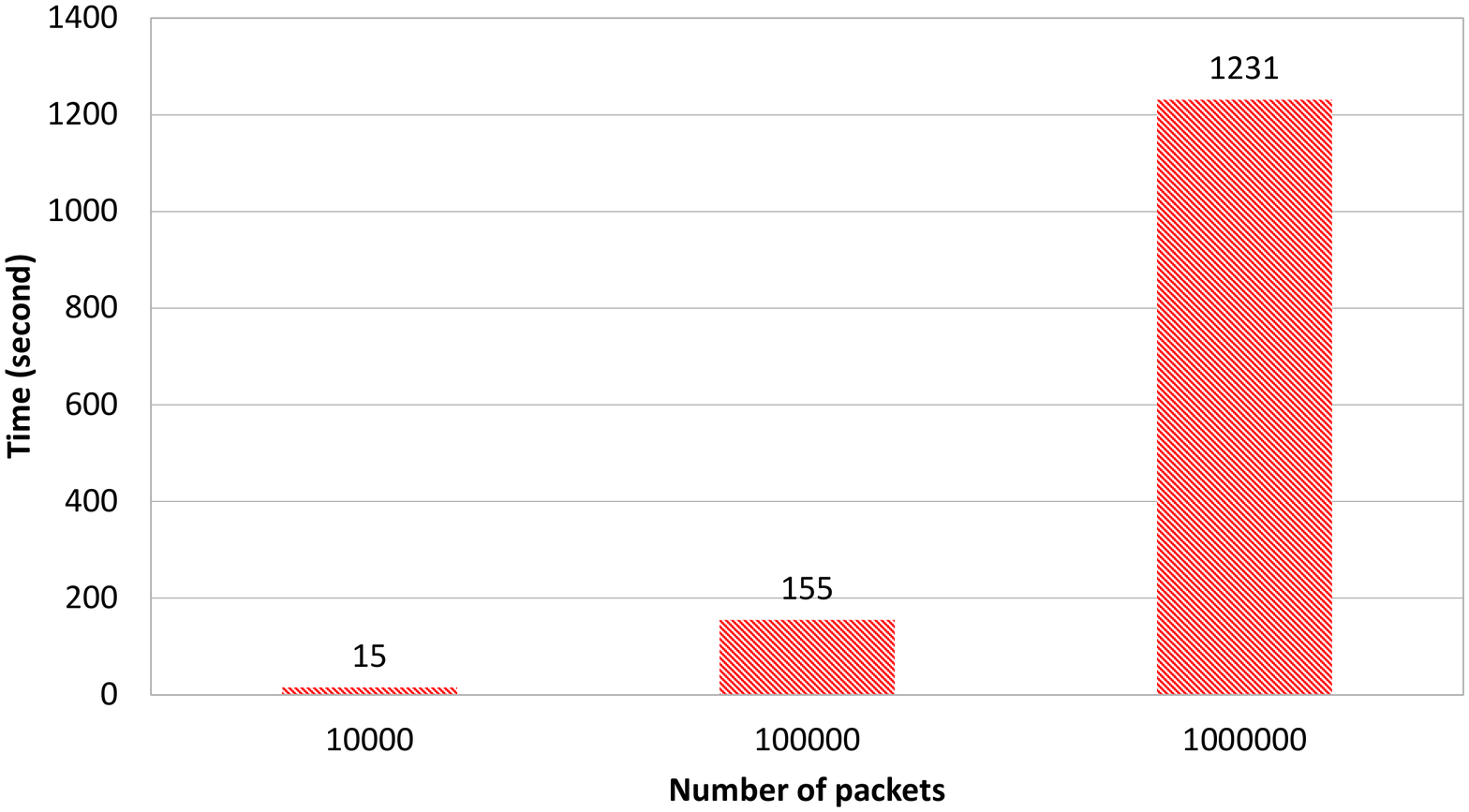}}
	\caption{Performance of creating \ac{DDoS} attack packets}
	\label{fig:id2t-packets-generating}
\end{figure}

\paragraph{b) \ac{DDoS} Attack Creation:} The toolkit provides attacks such as \ac{DDoS} attack, which usually produces a large number of packets. Therefore, the performance of generating packets is important for such toolkit. Figure~\ref{fig:id2t-packets-generating} shows the results of several performance experiments with regard to the time needed to generate $10^4$, $10^5$, and $10^6$ packets of a \ac{DDoS} attack. Although the experiments show that the time increases exponentially, it is illustrated that \ac{ID2T} is able to generate a big number of packets in a reasonable amount of time.

\section{Testing Intrusion Detection Datasets (TIDED)}\label{sec:tided}

Datasets are needed to accurately evaluate the detection capabilities of \acp{NIDS}. For evaluations to be accurate and unbiased, datasets need to satisfy the requirements presented in Section~\ref{sec:requirements-evaluation}. If these requirements are not satisfied, \acp{NIDS} might detect or learn artificial patterns that do not typically occur in networks. And while most requirements are easily verified, the \textit{Good Quality} non-functional requirement is of a more complex nature. In order to analyze datasets to identify potential \emph{quality} problems, we have developed the \ac{ID2T} module named \ac{TIDED}. The goal of \ac{TIDED} is to identify potential sources of \emph{artifacts} (synthetic defects, see Section~\ref{sec:artifacts}), inherent deficiencies of the background traffic (a user defined \Ac{PCAP} file), or both.

There are two sources of \emph{artifacts} or deficiencies that contribute to the fallibility of a dataset when evaluating \acp{NIDS}. The tools that inject synthetic attacks or manipulate traffic are the first source. The second source is the dataset itself and the environment where it is generated. \ac{ID2T} actively avoids creating \emph{artifacts} throughout the process of injecting synthetic attacks. In spite of these efforts, it is still possible that, after using \ac{ID2T}, a generated dataset is unsuitable for evaluation because of its own inherent or unforeseen issues.

In the past, many datasets have been found to contain inadvertent \emph{artifacts} or deficiencies. The well known DARPA 1999 dataset~\cite{cunningham1999evaluating}, for example, uses a limited set of \ac{TTL} values in all TCP headers. Unknowingly, attacks use \ac{TTL} values that background traffic does not use, making attacks easy to identify. The FLAME~\cite{Brauckhoff2008} tool injects synthetic attacks at the flow level. Each synthetic flow is injected with a predictable delay, making the synthetic flows easy to uncover. In Section~\ref{sec:artifacts}, a classification of \emph{artifacts} is presented with more examples like these.

\ac{TIDED} tests the reliability of any network capture file (in the \ac{PCAP} format). This module focuses on finding abnormal statistical properties that might point to \emph{artifacts} or defects that contradict the dataset's requirements. A set of statistical tests are run on top of a dataset. The results of these tests have a twofold use. On the one hand, they are used to measure and validate certain metrics belonging to the network's background traffic. On the other hand, the reliability tests are put at the disposal of \ac{ID2T}'s attack controller (see Figure~\ref{fig:architecture}). This enables the attacks to refine the parameters used by attacks to better replicate the network's background traffic. With this module, \ac{ID2T} becomes not only an attack injection tool, but also a network dataset analysis tool.

\subsection{Test Categories}\label{sec:tided-test-class}

\begin{figure}[h]
  \centering
  \includegraphics[width=0.8\linewidth]{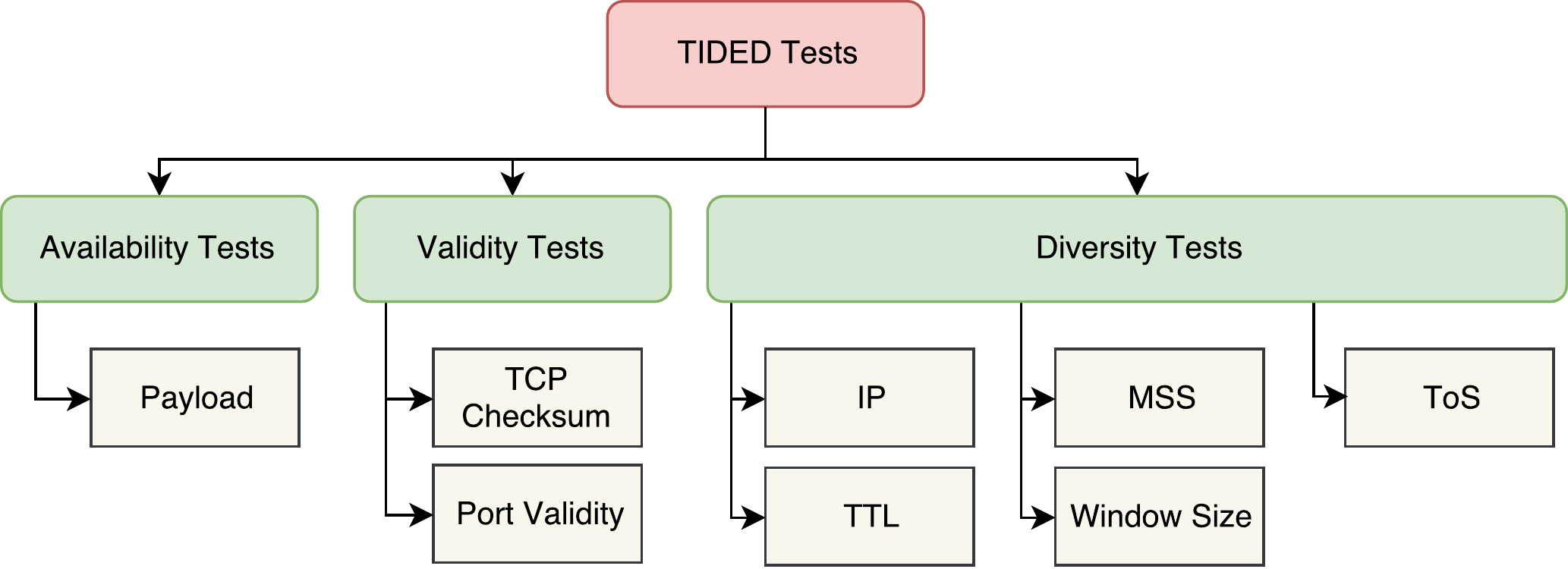}
  \caption{The categories into which the statistical tests performed by the TIDED module are divided.}
  \label{fig:tests_categories}
\end{figure}

The statistical tests performed by \ac{TIDED} can be categorized as shown in Figure~\ref{fig:tests_categories}. \emph{Availability Tests} refer to those that verify the availability of packets' payloads. \emph{Validity Tests} look for either TCP checksum problems, or invalid and uncommon ports. The port validity tests report the usage of standard but unassigned ports (in accordance to \ac{IANA}) and how many times port 0 is used. \emph{Diversity Tests} use a set of metrics (see Section~\ref{sec:divers-test-metr}) to present information related to the packets' header fields. With this set of metrics, we quantitatively qualify an analyzed network. This enables us to identify potential problems when certain qualities do not meet expectations. For example, if the analyzed traffic comes from a supposedly backbone network and the entropy of source IP addresses is below a threshold, there is evidence to suspect that the traffic is not from the claimed source. In the following, the tests are discussed in more detail.

\subsubsection{Availability Tests}

\paragraph{Payload availability} The ratio of packets with data payload against those without. A dataset without data payloads cannot be used by all \acp{NIDS}. Therefore, this information is needed to determine if a dataset is suitable only in restricted evaluation scenarios.

\subsubsection{Validity Tests}

\paragraph{TCP Checksums} The ratio of wrong TCP checksums against correct ones. This test helps to detect synthetic datasets that have not correctly generated the network packets. It can also identify networks with faulty hardware, tools with inadequate anonymizing mechanisms or packet capturing tools with deficiencies. In \cite{Mahoney2001}, for example, attacks could be identified based on incorrect packet checksums. We note that the complete absence of incorrect checksums is potentially an artifact or defect. Real traffic usually contains small quantities of network packets with incorrect checksums.

\paragraph{Port Validity} The counts of port numbers used in each of the three port ranges defined by \ac{IANA} in RFC 1340 \cite{rfc1340} and RFC 6335 \cite{rfc6335}. As a special case, the test reports the number of times packets targeted port zero. In the popular Berkeley sockets \ac{API}, port zero indicates that a random port should be utilized. In wrong implementations or incorrectly generated synthetic attacks, zero is used as a port. Real networks seldom observe packets directed towards port zero.

\subsubsection{Diversity Tests}

The following tests use various metrics, which will be explained on the next section.

\paragraph{IP Diversity} The goal here is to quantitatively characterize the diversity of source and destination IP addresses . The IP address diversity correlates with the network type (home, office, backbone, industrial control, etc.). A combination with the knowledge of the network type enables an analyst to identify potential issues.

\paragraph{TTL Diversity} This aims to characterize the diversity of the \ac{TTL} values of all network packets. Different sources alter the \ac{TTL} value of the packets, for example, different operating systems use different initial \ac{TTL} values and networking devices (e.g., routers, switches, etc.) alter these. The diversity of this field correlates with the diversity of the packets' sources and networking devices.

\paragraph{\ac{MSS} Diversity} The distribution of all observed \ac{MSS} values within a \ac{PCAP} file. Although \ac{MSS} values tend to remain constant in most packets, certain hosts choose to increase it to maximize their throughput. Conversely, a host may choose to lower the \ac{MSS} to reduce IP fragmentation. The distribution of the \ac{MSS} may correlate with a network's design (e.g., data or resource sharing, providing services, cloud computing, etc.).

\paragraph{Window Size Diversity} This shows the changing behavior of the window sizes of network packets. When troubleshooting network problems, the window size is often a key indicator to examine. Additionally, networks attached to different systems (e.g., workstations, clusters, grids, etc.), benefit from tuning the window sizes \cite{feng2003optimizing}.

\paragraph{ToS Diversity} This relates to the evolution of the \ac{ToS} TCP values. The meaning of the \ac{ToS} header field, originally defined in RFC791, has gone through five different definitions until its current definition, as defined in RFC2474. Therefore, the diversity of this field lets an analyst identify the year and version of the used TCP standard.

\subsection{Diversity Test Metrics}\label{sec:divers-test-metr}

With the help of a few metrics, an analyst may quantify and characterize key aspects of a network \ac{PCAP} file. ID2T is also able to use these metrics to create attacks that better replicate the input traffic. To express diversity, we use several metrics and graphs.

\subsubsection{(Shannon) Entropy distribution}

\begin{figure}[h]
  \centering
  \includegraphics[width=0.55\linewidth]{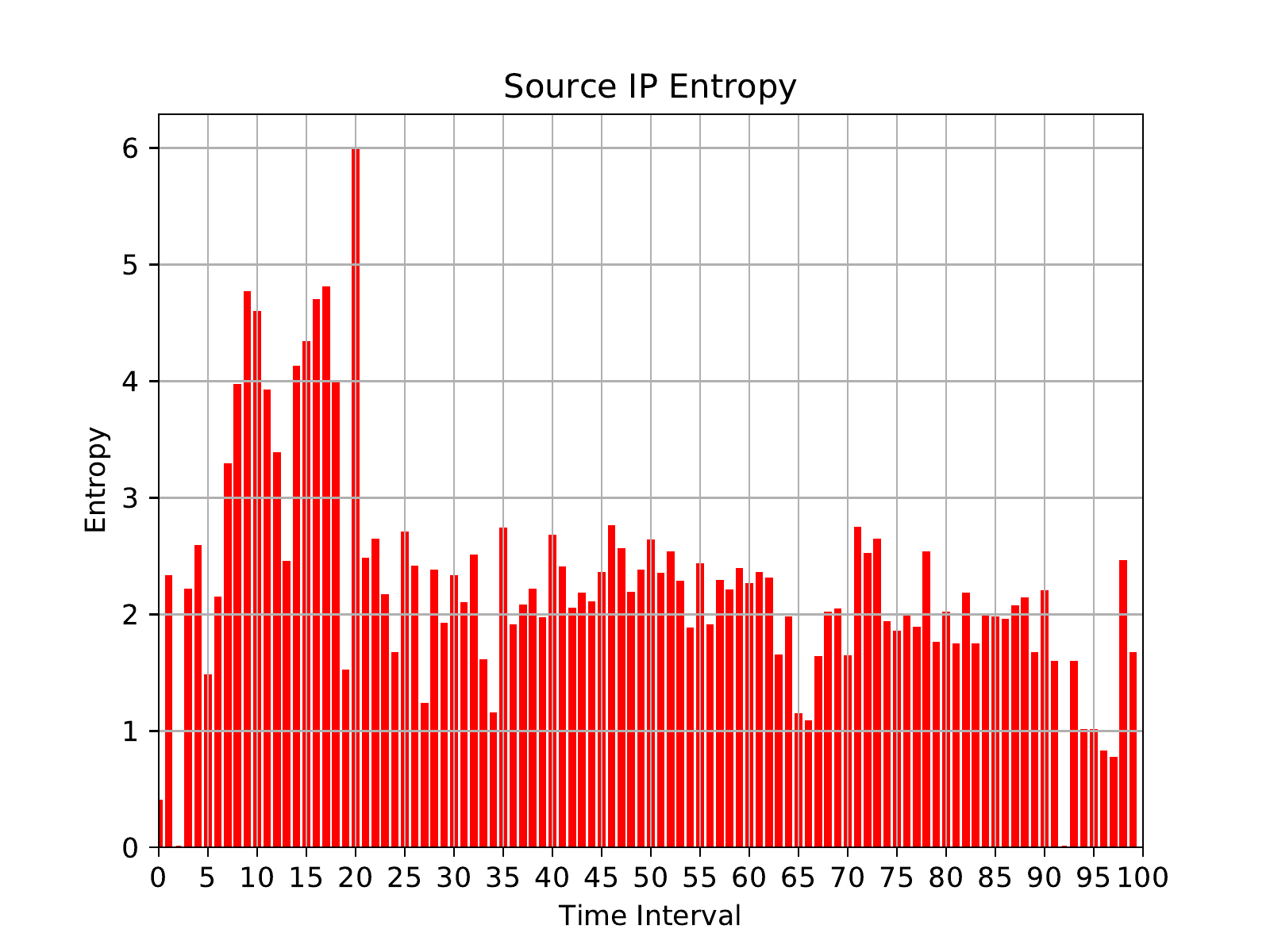}
  \caption{Example entropy graph of all the observed source IPs in a network PCAP file.}
  \label{fig:entropy_example}
\end{figure}

We use entropy to characterize the uncertainty of observing a particular feature within a time window. Entropy is calculated as:

\begin{equation}
H(X)=-\sum_{i=1}^{n} P(x_i) \cdot \log_2P(x_i),
\end{equation}

where \(X = \{x_1, x_2, \ldots, x_n\}\) and \(P(X)\) is the probability mass function of \(X\). The minimum entropy value is zero and the maximum is $\log_2n$, where $n$ is the total number of elements \(\Vert X \Vert\). Low entropy signals that some feature values rarely change within a time window; that is, the values of the feature are predictable. High entropy indicates that different values of the same feature are repeatedly seen; the predictability of such a feature is low. In Figure~\ref{fig:entropy_example}, we exemplify this metric by showing the entropy of the source IP addresses of all packets seen in a \ac{PCAP} file of the MAWI dataset (see Section~\ref{sec:mawi_dataset} for a description of the MAWI dataset). The file is analyzed in 100 intervals. It can be observed that in the first quarter of the capture, the traffic is highly irregular. Afterwards, the traffic settles and the observed source IP addresses are less randomly encountered.

\subsubsection{Normalized (Shannon) entropy}

The normalized entropy $H_n(X)$ is defined as the entropy $H(X)$ divided by the maximum entropy value $\log_2n$ for $n$ samples:
\begin{equation}
H_n(X) = - \sum_{i=1}^n \frac{P(x_i) \cdot \log_2 P(x_i)}{\log_2 n}.
\end{equation}

\begin{figure}[h]
  \includegraphics[width=0.65\linewidth,trim={1.9cm 3.3cm 2cm 3cm},clip]{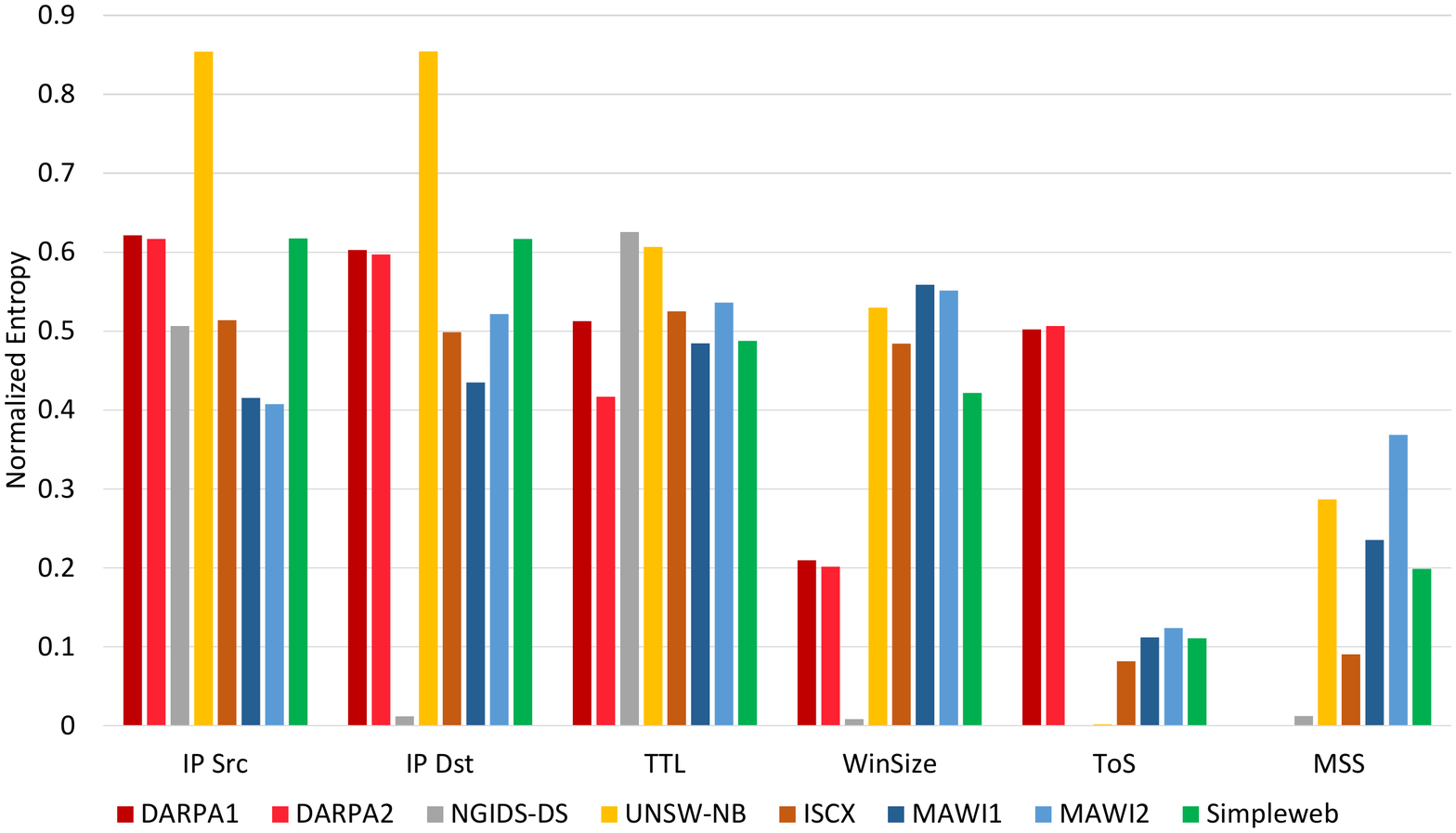}
  \caption{Comparison of normalized entropies between different datasets.}
  \label{fig:normalized_entropy_example}
\end{figure}

Normalized entropy lies in the range $[0, 1]$. Normalized entropy is used when comparing two random variables that have different number of samples $n$. Figure~\ref{fig:normalized_entropy_example} compares six different properties belonging to eight different datasets. The DARPA1 and DARPA2 bars correspond to two different \ac{PCAP} files found in the DARPA dataset. Similarly, the MAWI1 and MAWI2 bars corresponds to \ac{PCAP} file from different days of the MAWI dataset.

A few conclusions can be made from comparing the normalized entropy: The DARPA datasets have considerably high \ac{ToS} entropy which corresponds to the fact that DARPA was created when the \ac{ToS} field was defined to have a different meaning than more modern datasets. The DARPA dataset has deficiencies in how window sizes are chosen. When contrasting the normalized entropy of the window sizes, this deficiency of the DARPA dataset becomes obvious (the window sizes barely change due to how synthetic traffic was generated~\cite{Mahoney2003a}). The NGIDS-DS synthetic dataset also has some potential issues when looking at the \ac{MSS}, \ac{ToS} and window size. Furthermore, from the normalized entropy of the destination IP addresses, it can be seen that almost always the same hosts are contacted; signaling a potential lack of network diversity. From the same perspective of network diversity, the UNSW-NB almost always captures traffic of hosts that are only seen once. This is not an issue on itself. Depending on the purpose of the dataset, however, a high diversity might reduce its usability (e.g., learning patterns of normality for detecting anomalous host communications).

\subsubsection{Novelty distribution}

\begin{figure}[h]
  \centering
  \includegraphics[width=0.65\linewidth]{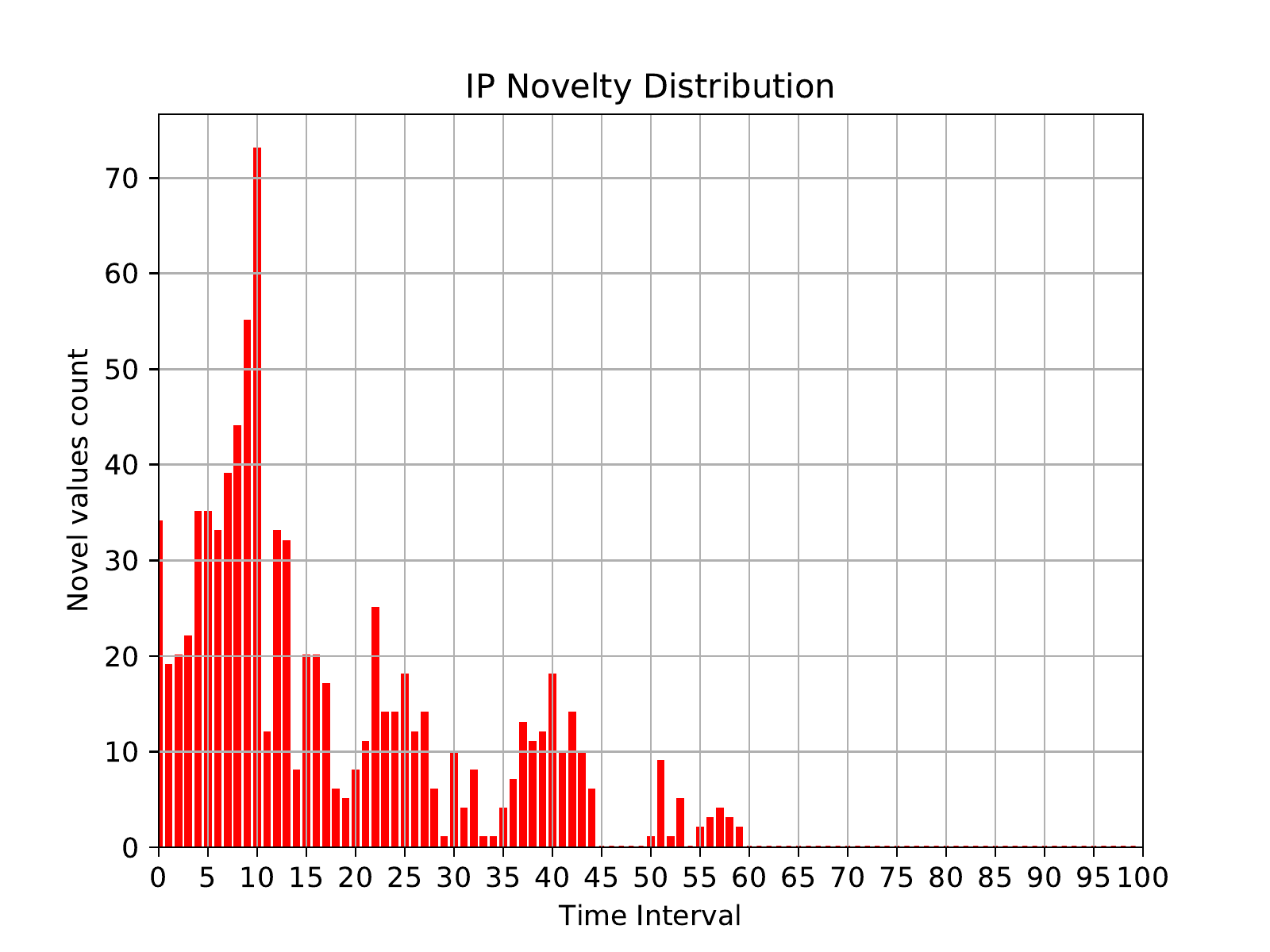}
  \caption{Novelty distribution of IPs of one day of the DARPA dataset. Halfway though the day, no records of new IP addresses are observed.}
  \label{fig:darpa_novelty_distribution}
\end{figure}

\begin{figure}[h]
  \centering
  \begin{subfigure}[h]{0.45\linewidth}
    \captionsetup{width=.8\linewidth}
    \includegraphics[width=\linewidth]{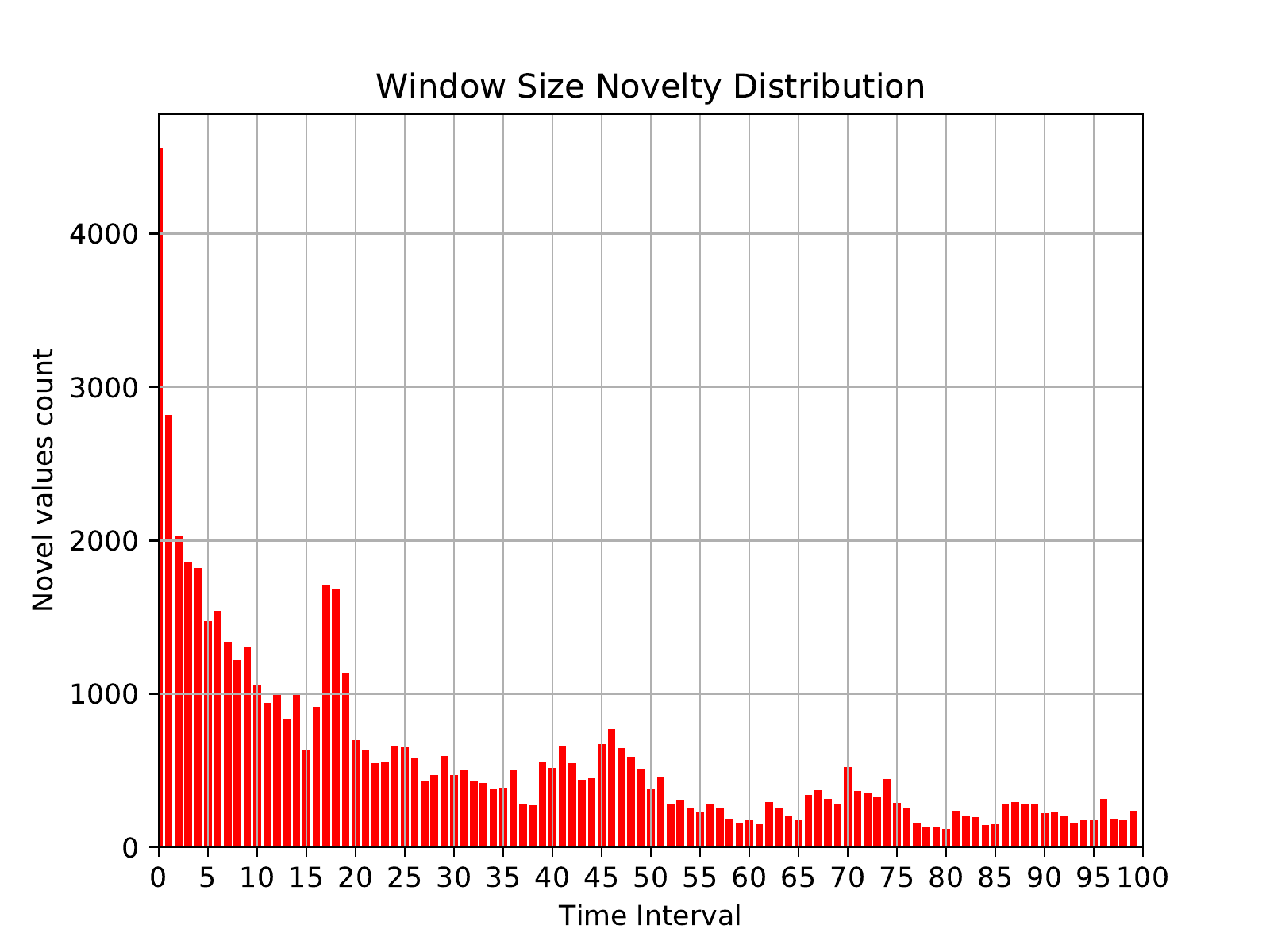}
    \caption{Novelty distribution of window sizes of a day of the MAWI dataset.}
    \label{fig:mawi_novelty_distrubtion}
  \end{subfigure}
  \begin{subfigure}[h]{0.45\linewidth}
    \captionsetup{width=.8\linewidth}
    \includegraphics[width=\linewidth]{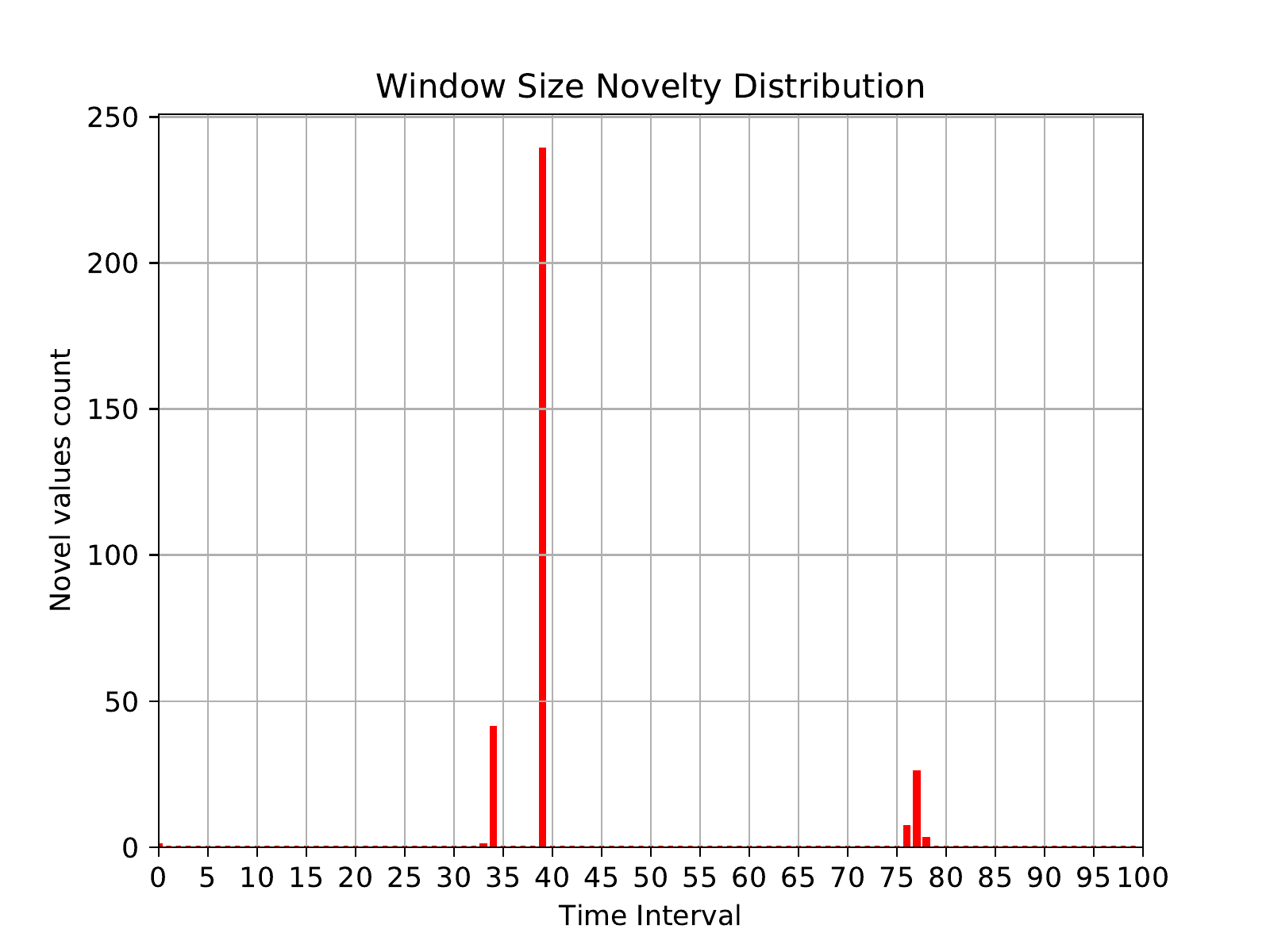}
    \caption{Novelty distribution of window sizes of a day of the UNSW dataset.}
    \label{fig:uswc_novelty_distrubtion}
  \end{subfigure}
  \caption{Example plots of the novelty distribution metric.}
  \label{fig:novelty_distribution_comparison}
\end{figure}

Distribution of newly observed values at different time windows. This metric partitions the dataset into time windows (x-axis). For each time window, the number of values (y-axis) that have never seen in the previous time windows are displayed. Figures~\ref{fig:darpa_novelty_distribution} and~\ref{fig:novelty_distribution_comparison} show examples of this metric for different network features.

In Figure~\ref{fig:darpa_novelty_distribution}, the distribution of novel IP addresses of a day of the DARPA dataset are shown. After time window 60, where each time window spans around 15 minutes, no new IP addresses are seen again.

In Figure~\ref{fig:mawi_novelty_distrubtion} and~\ref{fig:uswc_novelty_distrubtion}, the novelty distribution of window sizes of two datasets can be compared. In Figure~\ref{fig:mawi_novelty_distrubtion}, we see that new window sizes in one day of the MAWI dataset are always recorded (in diminishing counts). This is consistent with the fact that MAWI has data of a backbone network. In contrast, Figure~\ref{fig:uswc_novelty_distrubtion} shows that only in few time windows new values of window sizes are observed. This points to a potential dataset defect: this is not expected given that the UMSW-NB dataset is supposed to contain real network traffic with injected attacks for the evaluation of \ac{NIDS}.

\subsubsection{(Normalized) Entropy of novelty distribution}

The (normalized) entropy pertaining to a specific network feature as found in the entire dataset. Entropy is not only used for determining the predictability of a random variable, it is also used to characterize the shape of a distribution. This metric provides a quantitative measurement that provides a graphical representation as well as the means to compare different datasets.

\begin{figure}[h]
  \includegraphics[width=0.55\linewidth,trim={1.9cm 3.3cm 2cm 3cm},clip]{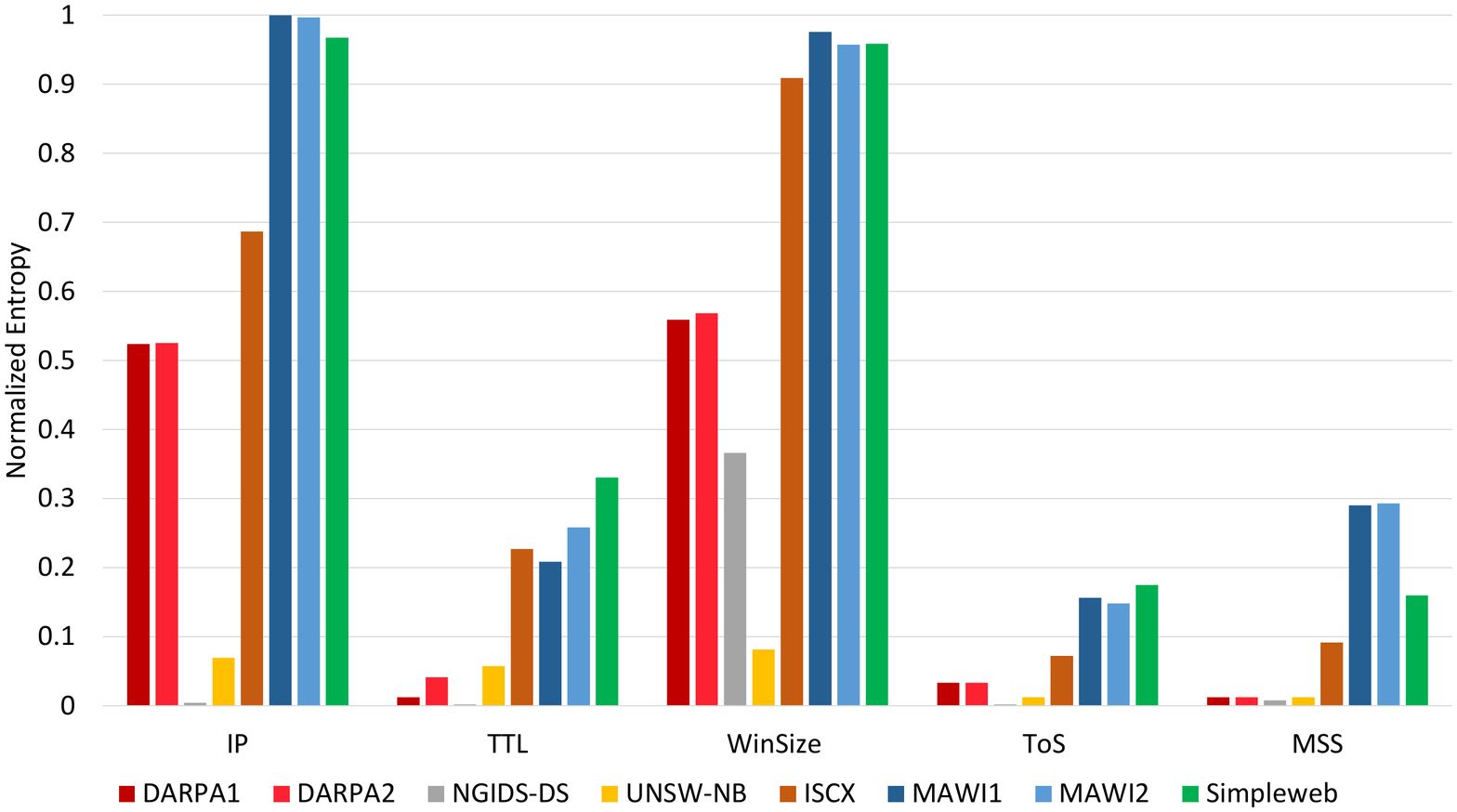}
  \caption{Comparison of the normalized entropies of the distribution of novel values between different datasets.}
  \label{fig:novelty_normalized_entropy_example}
\end{figure}

Figure~\ref{fig:novelty_normalized_entropy_example} shows a comparison of the normalized entropy of different network features as found in different datasets. From the figure, it is possible to see that the \ac{ToS}, \ac{TTL} and \ac{MSS} header fields of the DARPA dataset (DARPA1 and DARPA2) have seen a low number of novel values due to their low entropy. The same is true for the UNSW-NB dataset, with the addition that the window sizes and IP addresses have this same property. Due to the nature of the dataset, this is a potential issue. The MAWI (from two distinct days, MAWI1 and MAWI2) and Simpleweb datasets can also be assessed from the plot. Namely, these datasets record traffic from large networks and, therefore, never cease to observe new IP addresses (along with different window sizes, presumably due to congestion prevention mechanisms). 

\begin{figure}[h!]
  \centering
  \includegraphics[width=0.55\linewidth]{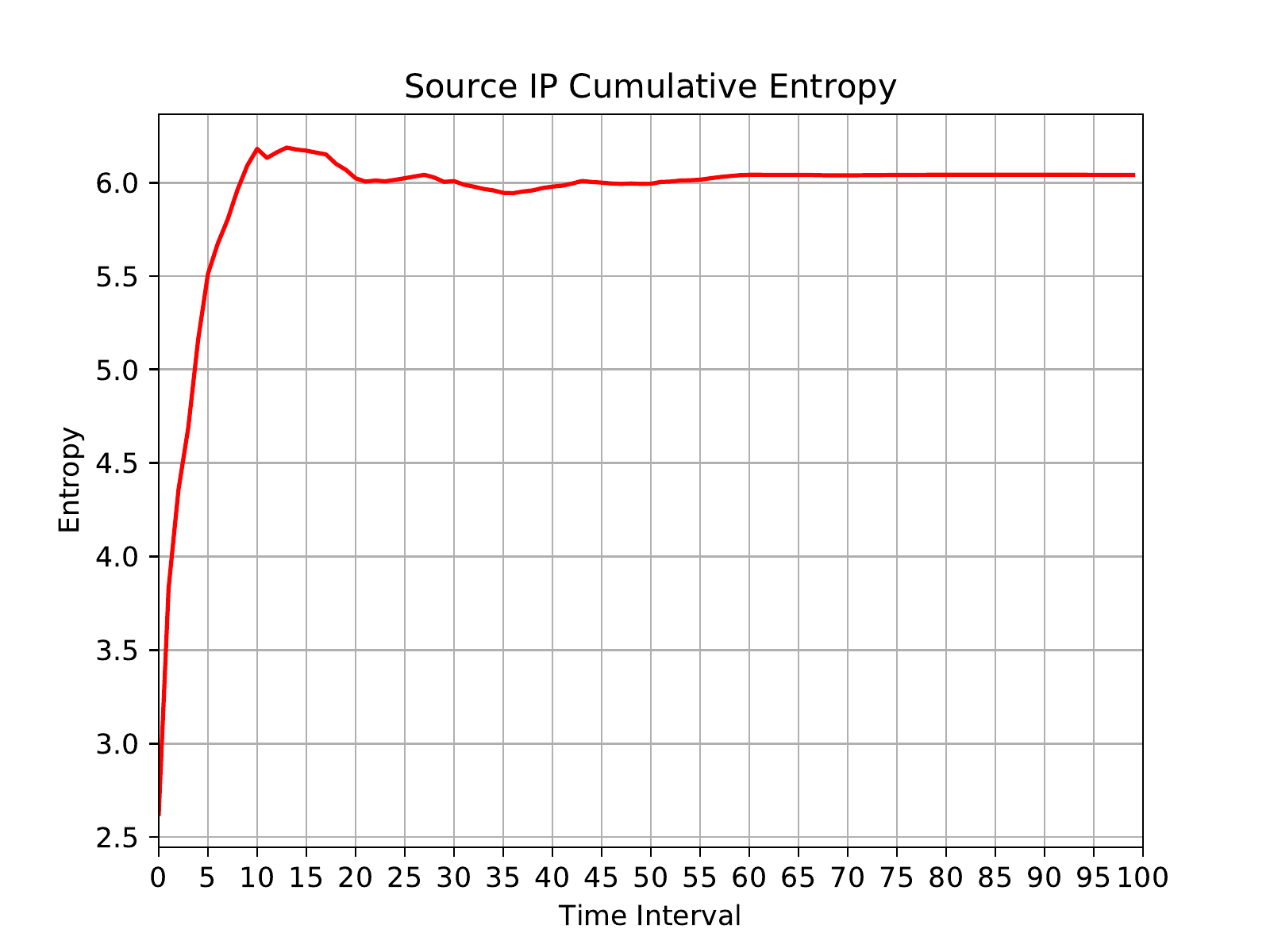}
  \caption{Cumulative entropy of a day in the DARPA dataset for the IP sources.}
  \label{fig:cumulative_entropy}
\end{figure}

\subsubsection{Cumulative (Shannon) entropy distribution}

The cumulative entropy of a dataset at different time windows. At each time window, all previous time windows are used to calculate a feature's entropy. This metric enables an analyst to identify unexpected entropy gains (or losses).

Figure~\ref{fig:cumulative_entropy} shows the source IP cumulative entropy of one day of the DARPA dataset. This plot is an alternative view of the entropy distribution plot (cf. Figure~\ref{fig:entropy_example}). In this example, we observe how no new source IP addresses are observed in one of the days of the DARPA dataset after time interval 60.


\section{Synthetic \ac{ID2T} Attacks}\label{sec:attacks}

\begin{figure}[ht!]
  \centering
  \includegraphics[width=0.70\linewidth]{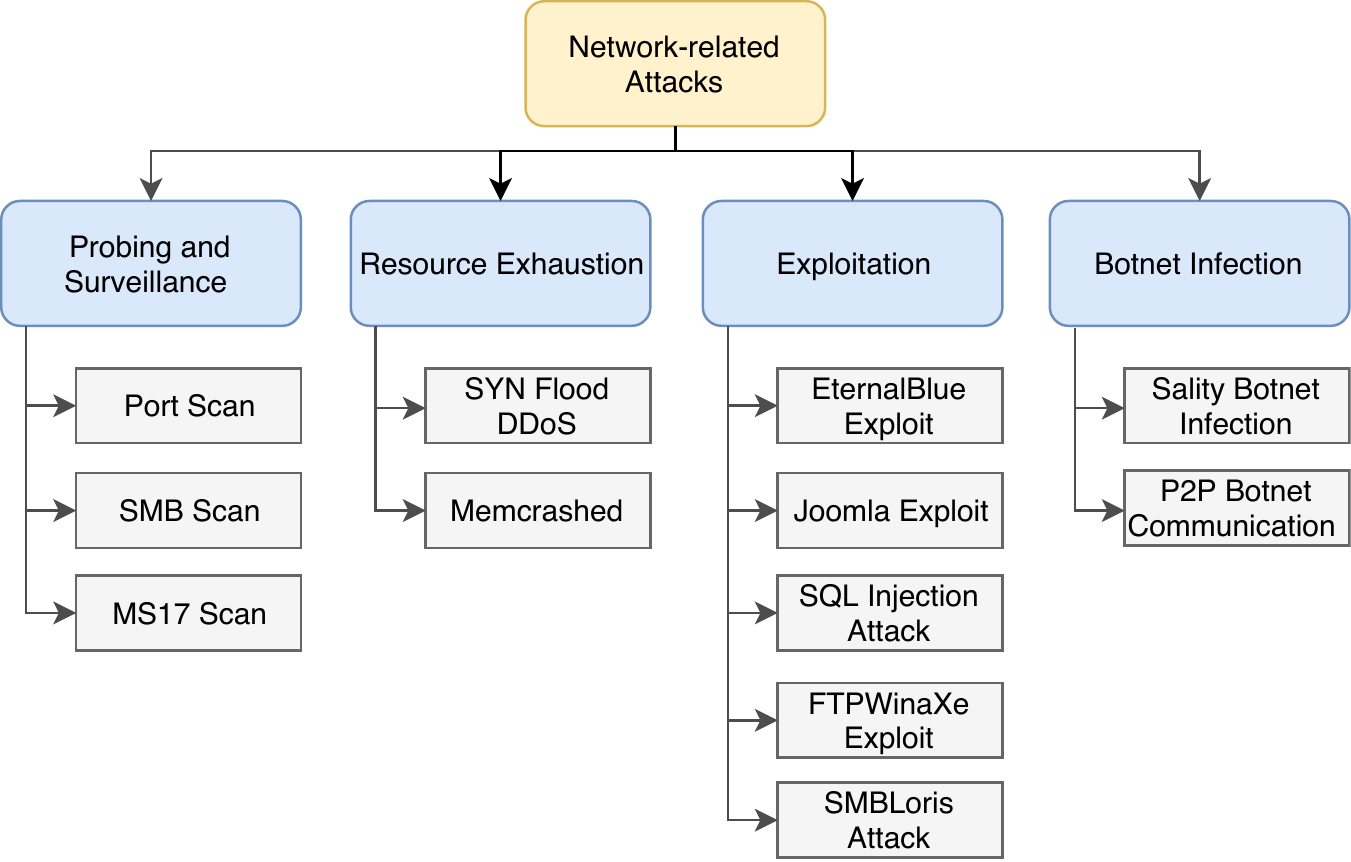}
  \caption{Classification and attacks provided by ID2T.}
  \label{fig:attacks_taxonomy}
\end{figure}

\ac{ID2T} puts several synthetic attacks at the disposal of the \ac{NIDS} community. Figure \ref{fig:attacks_taxonomy}, presents a classification of the attacks that are currently supported by \ac{ID2T}.
Each attack attempts to create synthetic traffic that replicates, whenever possible and desirable, the conditions of some background network traffic. For example, \ac{ID2T} replicates the distributions of various header fields' values, such as the \ac{TTL}, \ac{MSS}, and window size. Moreover, the rate of the injected packets is calculated in a way that considers the changes of the background traffic intensity, i.e., high traffic intensity leads to extra latency, thus, the packets are injected at a lower rate. Users can provide \ac{ID2T} with a set of parameters to adjust the generated attack. These parameters vary from one attack to another. If the user does not provide the required parameters, \ac{ID2T} uses the statistical properties of the background traffic to select proper values automatically. In addition, \ac{ID2T} considers the behavior of real-world tools, e.g., Nmap and Metasploit, that are commonly used to perform such attacks. In the following subsections, we will describe the categories of attacks in Figure \ref{fig:attacks_taxonomy} in more detail.

\subsection{Probe and Surveillance}\label{sec:probe-surveillance}
This class contains scan techniques which aim at collecting information about networks or hosts, usually, with the intention of preparing for further attacks. 
In \ac{ID2T}, three scans are provided as follows.

\subsubsection{Port Scan}\label{sec:portscan-attack}
\begin{figure}[ht!]
	\centering
	\begin{subfigure}[t]{0.5\textwidth}
		\centering
		\includegraphics[clip, trim= 2.5cm 3.5cm 3cm 3.5cm, width=\textwidth]{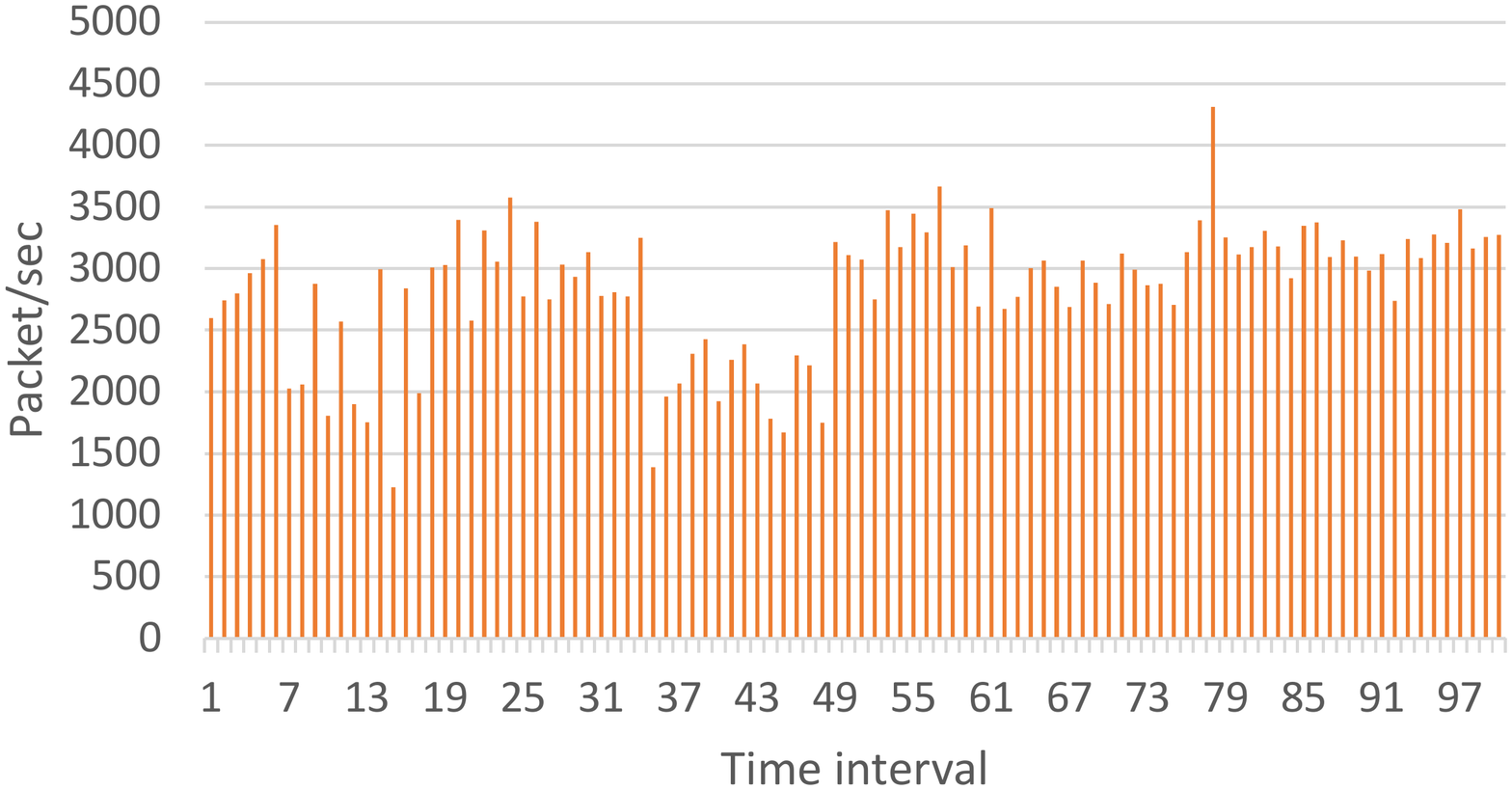}
		\caption{\label{fig:intervals_pkt_rate} Background traffic packet rate.}
	\end{subfigure}%
	~ 
	\begin{subfigure}[t]{0.5\textwidth}
		\centering
		\includegraphics[clip, trim= 2.5cm 9.5cm 2.5cm 10cm, width=\textwidth]{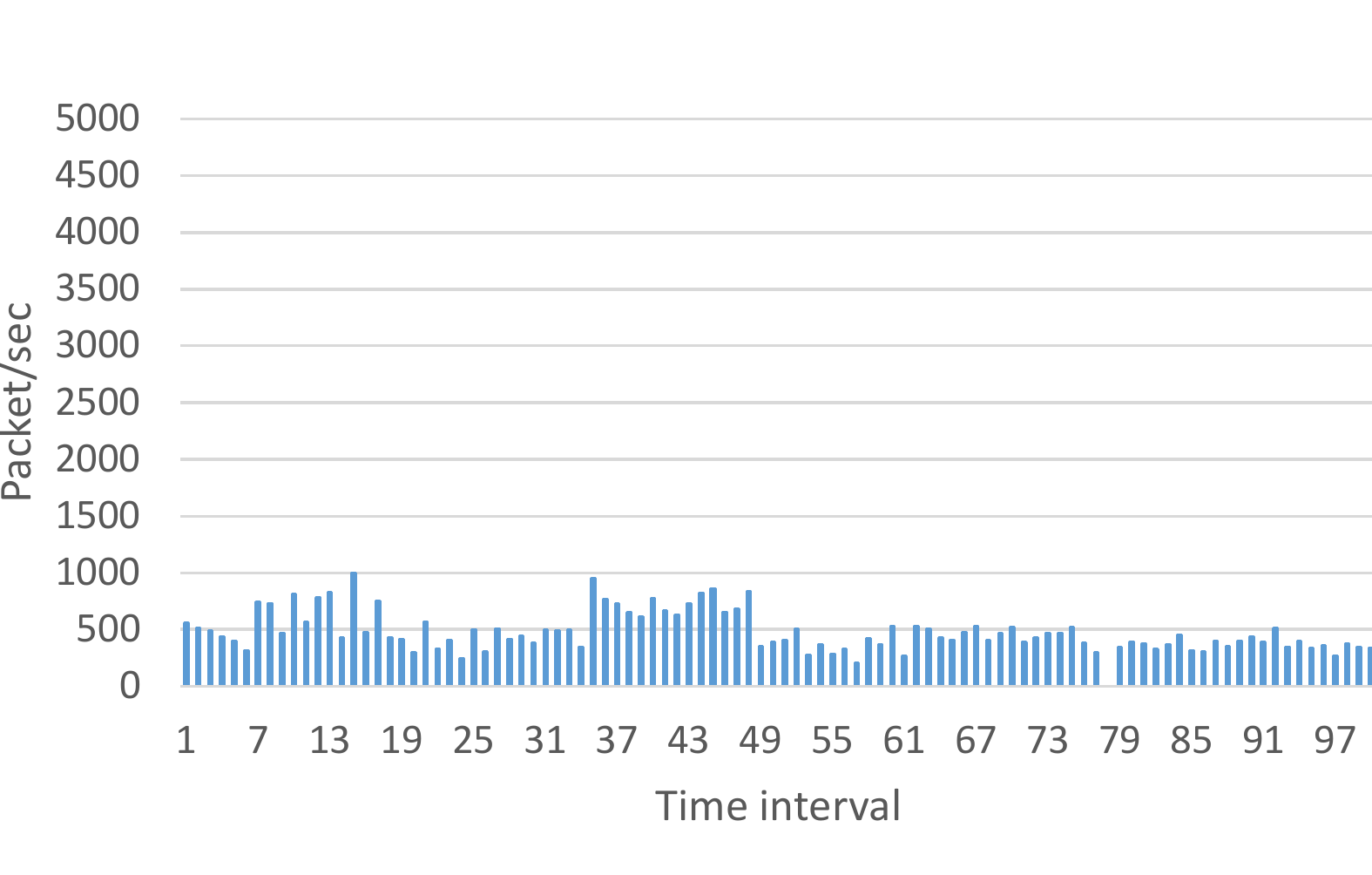}
		\caption{\label{fig:complement_pkt_rate}  Injected attack with a user-selected normalized packet rate.}
	\end{subfigure}
	\caption{\label{fig:mawi_pkt_rate} Packet rate of an arbitrary \ac{PCAP} file and an injected attack.}
\end{figure}

A port scan is a reconnaissance technique used to discover vulnerabilities in network hosts by sending port probes.
In \ac{ID2T}, a vertical \textit{TCP SYN} port scan is implemented. 
This scan targets various ports on a single host by sending \textit{SYN} packets.
\ac{ID2T} generates three types of packets, namely \textit{SYN}, \textit{SYN+ACK}, and \textit{RST}.
Furthermore, \ac{ID2T} imitates the behavior of an Nmap (default) scan with regard to the targeted ports; that is, the Nmap-service table is used to select the most common $1000$ open TCP ports. 
Nmap utilizes an adaptive packet rate technique, in which the packet rate corresponds to the changes of the background traffic intensity. \ac{ID2T} replicates this behavior. 
In particular, this was implemented by using the complementary packet rate of the background traffic after normalizing it to a user-selected value. 
Figure \ref{fig:intervals_pkt_rate}, shows the intervals of the packet rate of an arbitrary \ac{PCAP} file used as background traffic, while Figure \ref{fig:complement_pkt_rate}, depicts the complementary packet rate normalized to $1000$ packets per second, which is used for the injected packets rate. 

\subsubsection{SMB Scan}\label{sec:smbscan-attack}
This attack scans a network for \ac{SMB} servers. In particular, the attacker attempts to establish TCP connections with the victim(s) on the port 445. If the connection is established successfully, an \ac{SMB} negotiation starts and more packets are exchanged, where the attacker can learn more about the service provided by the victim (e.g., the version of \ac{SMB}).
\ac{ID2T} simulates this scan by generating and injecting the packets of TCP connections and \ac{SMB} negotiations.

\subsubsection{MS17 Scan}\label{sec:ms17scan-attack}
This specialized scan examines whether a victim has the MS17-010 patch. This patch resolves several vulnerabilities in the implementation of \ac{SMB}v1 in MS Windows OS. One of these vulnerabilities was used by the EternalBlue exploit. 
In Metasploit, the module \textit{smb\_ms17\_010} \footnote{https://github.com/rapid7/metasploit-framework/blob/master/modules/auxiliary/scanner/smb/smb\_ms17\_010.rb} performs this scan. \ac{ID2T} utilizes the packets of the \textit{smb\_ms17\_010} module as templates. It then injects these packets into the input \ac{PCAP} file after manipulating them to replicate the properties of the background traffic.

\subsubsection{Probing limitations}\label{sec:probe-limitation}
During probe activities, attackers receive responses from the victims and these responses are used to derive information about the victim. \ac{ID2T} provides a limited version of these responses. 
For example, in a real-world port scan, the response packets can be \textit{SYN+ACK}, \textit{RST}, or \textit{ICMP unreachable error} packets. In \ac{ID2T}, we consider one response packet type, which is the \textit{SYN+ACK} packet. However, detecting probe scans depends mainly on the traffic generated by the attackers. Therefore, the packets injected by \ac{ID2T} are sufficient to evaluate \acp{IDS} against these scans.

\subsection{Resource Exhaustion}\label{sec:resource-exhaustion}
The attacks in this class aim to use up the resources of networks or hosts, to deny legitimate users access a particular service. \ac{ID2T} provides two attacks under this class, namely, SYN Flood \ac{DDoS} and Memcrashed.

\subsubsection{SYN Flood \ac{DDoS} Attack}\label{sec:ddos-attack}
In this attack, a multitude of machines establish a massive amount of TCP connections with a victim to use up its resources. 
\ac{ID2T} creates this attack by generating two types of packets: \textit{SYN} packets, which are sent by the attackers, and \textit{SYN+ACK} packets, which are sent by the victim.
Known aspects from the victim previous activities in the background traffic are taken into account; first, to determine which (open) ports to attack, and second, to estimate the packet rate that is sufficient to bring the host down.
Moreover, \ac{ID2T} imitates some of the properties of Metasploit \ac{DoS} attack's packets; for instance, the random window size.

\subsubsection{Memcrashed Attack}\label{sec:memcrashed-attack} 
This is an amplification attack that exploits a vulnerability\footnote{CVE-2018-1000115} in Memcached\footnote{Distributed memory caching system used to speed up dynamic web applications.} servers, where an attacker sends forged UDP requests with a spoofed targeted source IP to servers. The servers send back responses to the targeted source IP, overwhelming its resources. Currently, \ac{ID2T} injects only the first part of this attack, i.e. the packets sent from the attacker to the Memcached servers.

\subsubsection{Resource exhaustion limitations}\label{sec:resource-limitation}
In real-world networks, resource exhaustion attacks leave usually remarkable impacts on the targeted networks and hosts, such as increasing the response time of the victim and creating network congestions, thus, increasing the network latency and causing packet loss. Replicating such impacts requires modification in the background traffic, e.g., by deleting packets, which is not considered currently in \ac{ID2T}.
However, the resource exhaustion attacks are usually recognizable by \acp{NIDS} based on the suspicious traffic generated by attackers, rather than the implications on the normal traffic. Thus, \ac{ID2T} datasets can be used effectively to evaluate \acp{NIDS} against these attacks.

\subsection{Exploitation}\label{sec:privilege-escalation}
These attacks target an existing bug or vulnerability in a system with the intention of gaining control, privilege escalation, or denying a service. Five different exploits are available in \ac{ID2T}. 

\subsubsection{EternalBlue Exploit}\label{sec:eternalblue-exploit}
This exploit targets a buffer overflow vulnerability\footnote{CVE-2017-0144} in the \ac{SMB}v1 in MS Windows OS. The Metasploit module \textit{eternalblue\_doublepulsar}\footnote{https://github.com/ElevenPaths/Eternalblue-Doublepulsar-Metasploit} performs this attack. \ac{ID2T} injects the packets generated by this module after manipulating the header fields, while maintaining the payload since it contains the malicious code. During this attack, several TCP connections are established. \ac{ID2T} takes into account preserving the conditions of these connections with regard to the order, overlap, and dependency.

\subsubsection{FTPWinaXe Exploit}\label{sec:ftpwinaxe-exploit}
In this attack, a malicious FTP server sends packets with overly long payloads to a WinaXe 7.7 \footnote{An X Windows environment, enables different OSs and their applications to be connected through SSH, TCP/IP, NFS, FTP, TFTP and Telnet.} FTP client, exploiting a buffer overflow vulnerability. A user can provide the payloads as input, otherwise \ac{ID2T} generates random ones.

\subsubsection{Joomla Privilege Escalation Exploit}\label{sec:joomla-priv-escal}
This exploit uses a vulnerability\footnote{CVE-2016-8870} found in Joomla\footnote{Content management system for web applications.} versions 3.4.4 through 3.6.3. The vulnerability allows attackers to create an arbitrary account with administrative privileges. \ac{ID2T} uses template packets obtained from the Metasploit \textit{joomla\_registration \_privesc} module\footnote{https://www.rapid7.com/db/modules/auxiliary/admin/http/joomla\_registration\_privesc}. In this attack, \ac{ID2T} manipulates the packets' header fields and the HTTP headers of the payload.

\subsubsection{SMBLoris Attack}\label{sec:smbloris-attack}
This attack exploits a vulnerability in the \ac{SMB} protocol that allows an attacker to make large memory allocations without being authenticated. \ac{ID2T} creates this attack by targeting the \ac{SMB} port of the victim with \ac{NBT} packets that have the maximum value in the length field.

\subsubsection{SQL Injection Attack}\label{sec:sql-injection-attack}
This attack targets a vulnerability\footnote{CVE-2016-2555} found in ATutor 2.2.1 \footnote{Content management system for education purposes.} applications. The vulnerability allows attackers to inject SQL statements, bypass authentication, and gain administrator privileges. Metasploit \textit{atutor\_sqli} module performs this attack. \ac{ID2T} injects the packets of this module after modifying them to adapt with the background traffic characteristics and the user parameters.

\subsubsection{Exploitation limitations}\label{sec:exploit-limitation}
This class of attacks are mainly distinguished by the packet payload, where the exploit is located. \ac{ID2T} effectively mimics such attacks by copying real malicious payloads. 
However, \ac{ID2T} is limited to producing specific versions of these attacks. For example, in real-world EternalBlue, the number of connections can vary based on the victim's resources, while in \ac{ID2T}, a fixed number of connections is generated.

\subsection{Botnet Infection}\label{sec:botnet-infection}
A botnet is a set of network-connected compromised machines that work in a coordinated fashion for malicious purposes, such as email spam delivery and performing \ac{DDoS} attacks. \ac{ID2T} provides the ability to inject two types of botnet traffic: a variant of Sality botnet and user-defined botnet communication patterns.

\subsubsection{Sality Botnet}\label{sec:sality-botnet}
Sality is a classification of malicious programs that infect executable files in MS Windows OS. 
Over time, Sality programs were developed to contain a variety of abilities, such as exfiltrating sensitive data.
The Sality variant that is supported in \ac{ID2T} is known as \textit{Win32-Sality.AM}. It loads a malicious DLL file in the memory of the infected host.
\ac{ID2T} injects the traffic of this botnet, which was obtained from \textit{VirusTotal.com}, after modifying the packet headers and the packet rate.

\subsubsection{P2P Botnet Communication}\label{sec:membersmgmt-botnet}
\ac{ID2T} is also able to generate P2P botnet communication traffic. A user needs to provide \ac{ID2T} with a specification of the attack, namely, a \ac{CSV} file, where the interactions and type of messages exchanged between bots are specified. 
The \textit{Attacks} module in \ac{ID2T} handles this file and generates the botnet traffic based on its content.
\ac{ID2T} can either add this traffic to existing hosts in the input \ac{PCAP} or generate new hosts acting as the bots. 

\subsubsection{Botnet infection limitations}\label{sec:botnet-limitation}
In the \ac{ID2T} version of the Sality botnet, a fixed set of malicious IP addresses is currently used. 
In real-world, these addresses can be more diverse and can be changed dynamically. 
However, \acp{NIDS} can detect the Sality botnet traffic not just based on the malicious IP addresses, but also based on footprints in the HTTP headers, which are replicated by \ac{ID2T}.


\section{Use Cases: Applications of ID2T}%
\label{sec:evaluation}

In this chapter, we demonstrate how ID2T can be applied to determine the detection capabilities of different \Acp{IDS} using replicable datasets. We illustrate this with two use cases. In the first use case, we inject attacks into a publicly available \Ac{PCAP} file and use an anomaly detection \Ac{IDS} to demonstrate the detection capabilities of the system. In the second use case, we use publicly available signature-based \Acp{IDS} to demonstrate that the synthetic attacks injected by ID2T are indeed triggering expected signatures. Therefore, assessing whether the signature-based \Ac{IDS} is working as intended. These use cases show how ID2T can be used to compare or test existing systems with datasets that can be replicated.

\subsection{Assessing Anomaly Detection Capabilities}%
\label{sec:ident-attacks-using}

As a first use case, we use \Ac{ID2T} to demonstrate the detection capabilities of an anomaly-based \Ac{NIDS} in a backbone network scenario. For such demonstration, we need a labeled \Ac{PCAP} of backbone traffic that contains the attack we wish to detect. The MAWI dataset~\cite{Fontugne2010} provides good candidate \Acp{PCAP} that include labels. Nonetheless, the provided labels do not reflect the ground truth concerning the network traffic stored in the \Acp{PCAP}. The labels mark anomalous traffic in accordance to three different anomaly detectors. To determine the type of attacks, a heuristic is used. Because no human is involved in the labeling process, the accuracy of the labels cannot be guaranteed. Furthermore, due to the amount of network traffic, it is also unfeasible to assess if there are other unlabeled attacks. With ID2T, we can inject precise attacks into sanitized MAWI \Acp{PCAP} files (e.g.\ removing traffic labeled as anomalous) to determine if an \Ac{IDS} can at least detect the injected attacks. This approach enables us to test the detection accuracy (and precision) of an \Ac{IDS} against a specific set of chosen attacks.

We demonstrate the detection capabilities of an anomaly detection methodology that uses shallow autoencoders, also known as \Acp{RNN}~\cite{cordero16rnn}. The demonstration, however, is applicable to any other anomaly-based \Ac{NIDS}. \Acp{RNN} learn how the entropies of traffic features behave within a time window. When testing for anomalies within a time window, the \Ac{RNN} assigns an anomaly score to the network traffic based on what it learned. It is assumed that traffic that does not conform to the learned behavior will yield higher anomaly scores in contrast to benign traffic. When the anomaly score passes a threshold determined by the model, an anomaly is raised (signaling a potential attack).

We use four days (from 2018/04/01 to 2018/04/04) of backbone network traffic of the MAWI dataset to train, inject and detect port scans and \Ac{DDoS} attacks using an \Ac{RNN}. We choose these two attacks as these are expected to leave distinguishable network fingerprints that an anomaly detector should identify. Each day of the MAWI dataset consists of 15 minutes of traffic. We processed each day using a three step process. First, we split each day in time windows of 10 seconds. Second, we convert all traffic within a time window into network flows. The resulting flows are sanitized by removing those marked as anomalous by the labels provided by the MAWI dataset. Finally, we calculate the entropy of source and destination IPs and ports (for a total of four entropies) of all flows (see \cite{cordero16rnn} for more details). The set of entropies belonging to the time windows of the first three days are used to train an \Ac{RNN}. The last day is used for testing purposes.

\begin{figure}[h]
  \centering
  \includegraphics[width=0.6\linewidth]{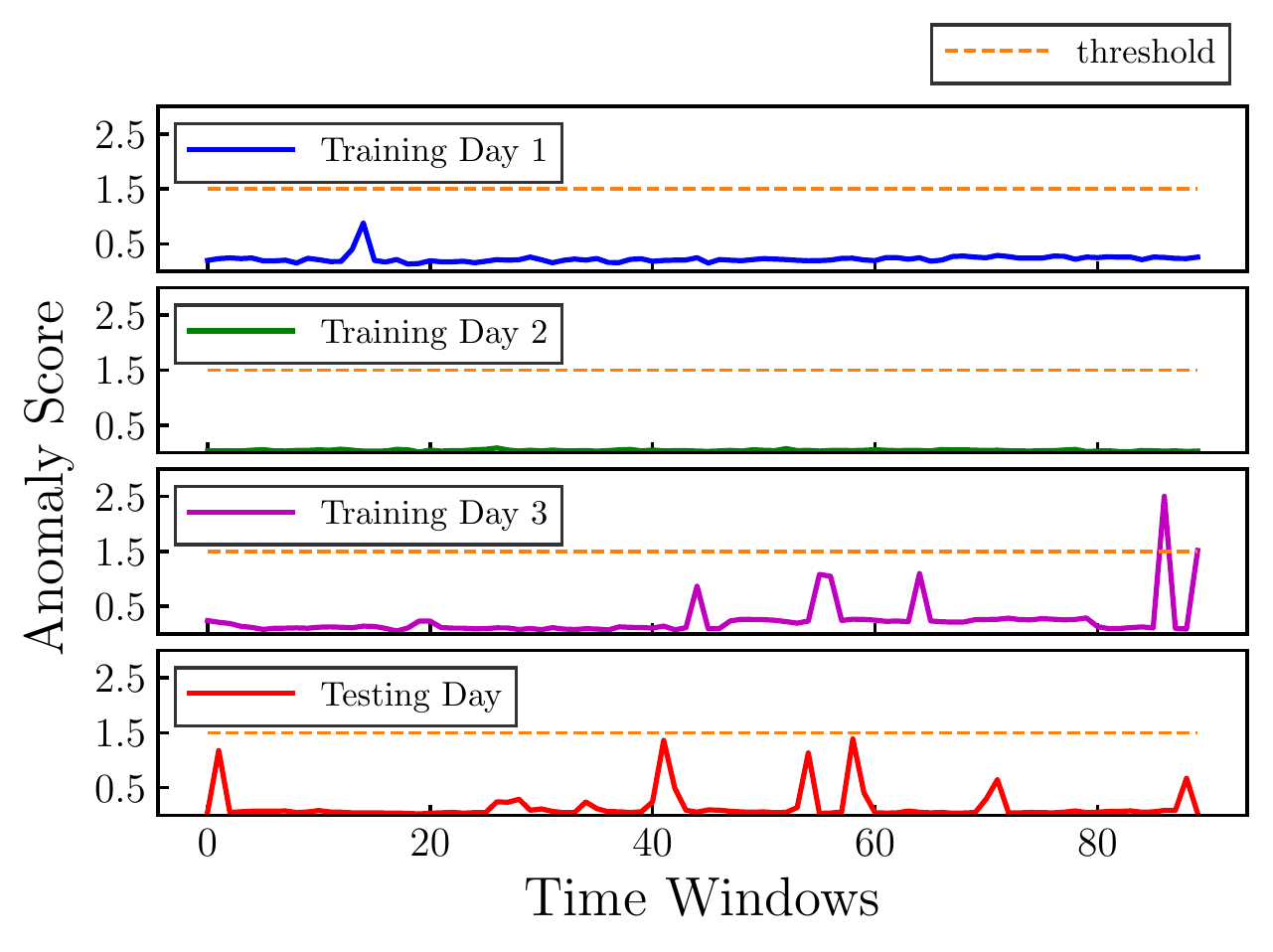}
  \caption{Anomaly scores given to four different days of the MAWILab dataset by an RNN. The first three days are used to train the model. The fourth day is used to test the trained model.}
  \label{fig:anomaly_scores}
\end{figure}

Figure~\ref{fig:anomaly_scores} shows the anomaly scores of the training and testing days as given by the trained \Ac{RNN}. The x-axis shows the different time windows into which a single day is split (15 minutes \(\times\) 60 seconds = 90 time windows). The y-axis shows the anomaly score of the entropies calculated at each time window. The \Ac{RNN} uses the first three days for training and is, therefore, expected to yield low anomaly scores for these training days. This is generally the case except for two instances within the third day (at time window 85 and 90). After manually examining those time windows, we identified that both spikes in the anomaly score correspond to an irregular amount of observed source ports (the entropy of the source ports more than doubled). By removing the outliers, the mechanism established an anomaly score threshold of 1.5. That is, any time window with an anomaly score above the dashed line is considered to have experienced anomalous traffic.

\subsubsection{Detecting \Ac{DDoS} Attacks}%
\label{sec:detect-ddos-attacks}

\begin{figure}[h]
  \centering
  \includegraphics[width=\linewidth]{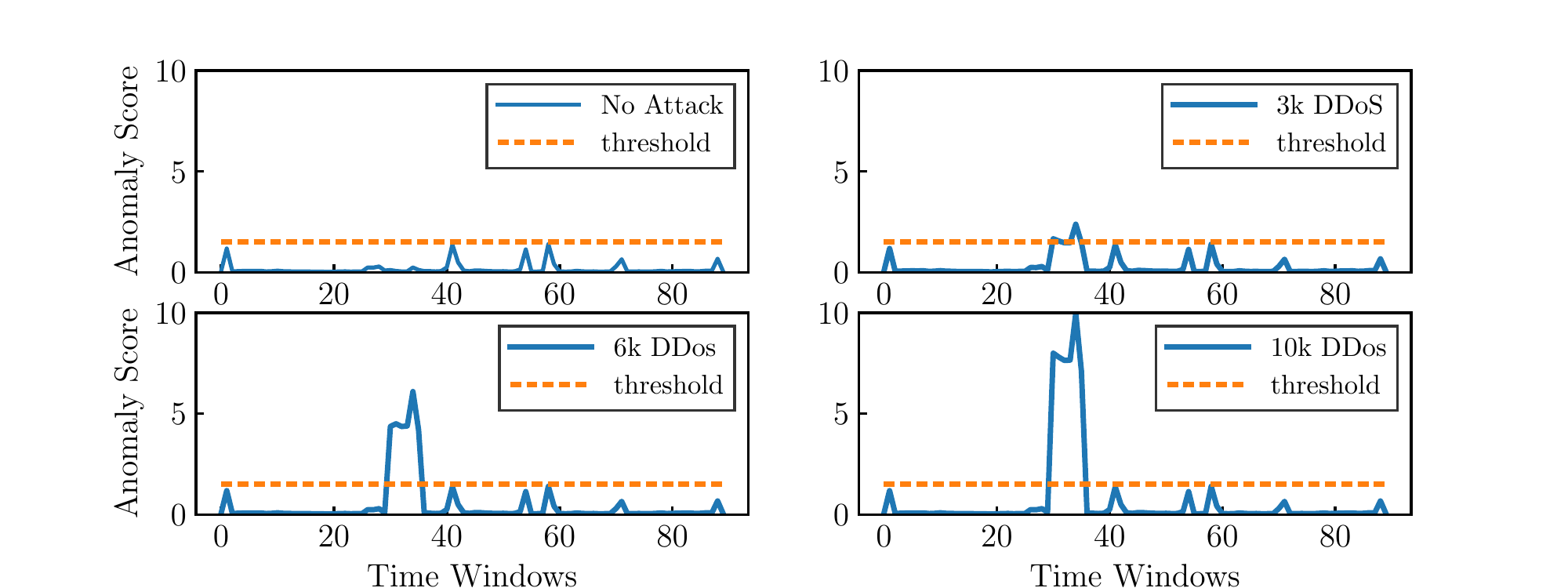}
  \caption{Detecting different \Ac{DDoS} attacks injected with ID2T using anomaly detection.}%
  \label{fig:ddos_attack}
\end{figure}

We proceed to inject \Ac{DDoS} attacks (see Section~\ref{sec:ddos-attack}) of different intensities into the testing day and use a \Ac{RNN} to detect the attacks. Figure~\ref{fig:ddos_attack} shows the resulting anomaly scores of the testing day with three different \Ac{DDoS} intensities. All attacks are injected into a clean testing day (attacks are not stacked) at time window 30. All attacks last a total of one minute (until time window 36). The upper left plot, shows the anomaly scores of the testing day with no injected attacks for reference. The upper right plot shows a 3,000 packets per second \Ac{DDoS} attack. The bottom left plot shows a 6,000 packets per second \Ac{DDoS} attack. The plot to the bottom right shows a high intensity \Ac{DDoS} attack of 10,000 packets per second. The \Ac{PCAP} file of the testing day contains 78 million packets. The percentage of packets added to the \Ac{PCAP} is, from the low to high intensity attacks, 0.23, 0.46 and 0.77 percent, respectively.

From Figure~\ref{fig:ddos_attack}, we observe that the injected \Ac{DDoS} attacks are detected. The anomaly scores of the time windows where the attacks lie (between 30 and 36) are above the predefined threshold. With the help of \Ac{ID2T}, we are able to demonstrate the \Ac{DDoS} detection capabilities of \Acp{RNN}. This use case scenario shows how \Ac{ID2T} can be used to reproduce and replicate the conclusions found in other publications without necessarily having to use the same dataset. In this case, we use the same type of background traffic and attack. The original \Ac{RNN} publication, however, uses different background traffic (days of the MAWI dataset from almost two year ago in relation to this work).

\subsubsection{Detecting Port Scans}%
\label{sec:detecting-smb-scans}

\begin{figure}[h]
  \centering
  \includegraphics[width=\linewidth]{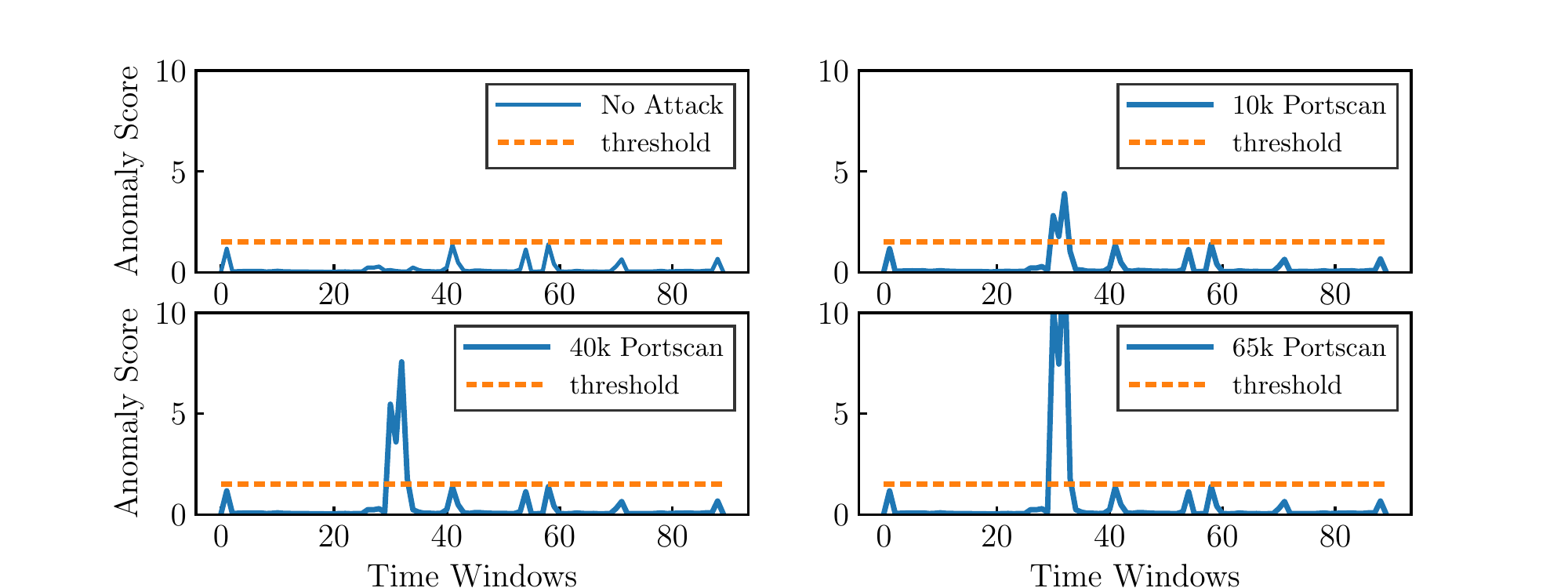}
  \caption{Anomaly scores of different port scans of varying intensities.}%
  \label{fig:portscan_scan}
\end{figure}

Using ID2T, we test the capability of \Acp{RNN} to detect port scans in large networks. Figure~\ref{fig:portscan_scan} shows plots of the anomaly scores of three different datasets injected with different port scan intensities. All injections take place at time window 30 and last 30 seconds (3 time windows). On the top left of the figure, as a reference, we show the anomaly scores of the target dataset without injected attacks. On the top right, we show the anomaly scores of the dataset after injecting a port scan targeting 10,000 random ports of five different host. On the bottom left, we show the anomaly scores when the injected port scan targeted five hosts and randomly scanned 40,000 ports. Finally, on the bottom right, we show the anomaly scores when the port scan targeted five hosts; scanning 10,000 random ports. The \Acp{RNN} are able to detect all port scans as can be seen by the anomaly scores that go beyond the marked threshold.

\subsection{Testing Configurations of Signature-based NIDSs}%
\label{sec:ident-attacks-using-1}

In the second use case, we test the configuration of two signature-based \Acp{NIDS} using \Ac{ID2T}. By injecting \Ac{PCAP} files with specifically chosen exploits or malware, we test an arbitrary configuration of Bro\cite{paxson1999bro} and Suricata\footnote{https://www.openinfosecfoundation.org}.

\newcolumntype{Y}{>{\centering\arraybackslash}X}
\newcolumntype{Z}{>{\centering\arraybackslash\hsize=.5\hsize}X}
\begin{table}[h]
  \centering
  \begin{threeparttable}[b]
    \begin{tabularx}{0.85\textwidth}{YZZ}
\toprule
 & Bro & Suricata\\
\midrule
Port Scan & \textcolor{olive}{\checkmark}\tnote{1} & \textcolor{red}{\sffamily x}\\
EternalBlue Exploit & \textcolor{olive}{\checkmark} & \textcolor{olive}{\checkmark}\tnote{2}\\
FTP WinaXe Exploit & \textcolor{red}{\sffamily x} & \textcolor{red}{\sffamily x}\\
Sality Botnet Infection & \textcolor{olive}{\checkmark} & \textcolor{olive}{\checkmark}\tnote{3}\\
\bottomrule
    \end{tabularx}
    \begin{tablenotes}
      \scriptsize
      \item [1] Identified as: Scan::Port\_Scan 188.165.214.141 scanned at least 15 unique ports of host 188.165.214.253 in 0m4s.
      \item [2] Exploit identified as: ETERNALBLUE MS17-010 Echo Reponse and ETPRO TROJAN (possible Metasploit payload).
      \item [3] Sality identified as: ETPRO TROJAN Win32/Sality.AM and ET MALWARE Sality Virus User Agent Detected (KUKU)
    \end{tablenotes}
    \caption{Testing different IDSs againt four different attacks injected by ID2T.}%
    \label{tab:ids-comparison}
  \end{threeparttable}
\end{table}

Using ID2T, we inject four different attacks into a \Ac{PCAP} file of office network traffic collected during a one minute interval. One minute of captured network traffic was close to 30 MiB of data and 55,000 packets. Because we inject small specialized attacks (that generate no more than 500 packets), this gives us enough room to hide our attacks. Table~\ref{tab:ids-comparison} shows the detection results of Bro and Suricata when only basic signatures are installed. We mark successful detection with an arrow (\textcolor{olive}{\checkmark}). A cross (\textcolor{red}{\sffamily x}) indicates that the corresponding malicious activity was not detected. We emphasize that these are not the results of testing \Acp{NIDS} themselves, but rather an arbitrary set of signatures we installed. The port scan was detected by Bro. One of Bro's anomaly rules found that the IP that we specified as an attacker was conducting port scans. Suricata, on the other hand, was not configured with signatures capable of detecting port scans. Both \Acp{NIDS} are able to detect the popular ``EnternalBlue'' exploit. Suricata detected, additionally, that the payload being used potentially originated from Metasploit. \Ac{ID2T} mimics what Metasploit does to create the ``EternalBlue'' exploit. The FTP WinaXe Exploit is a difficult attack to detect. A signature for it cannot be easily created as the exploit relies on overflowing a particular FTP command that only affects some vulnerable versions of the WinaXe program. Hence, as to be expected no \Ac{NIDS} detected the malicious payload. Finally, both \Acp{NIDS} correctly identify a Sality bot infection. Suricata is also able to detect that the user agent of some HTTP traffic is made by Sality. ID2T mimics how Sality sends HTTP messages with the user agent field set to ``KUKU''.

\subsection{Discussion of the Use Cases}%
\label{sec:discussion}

The previous two use case scenarios help illustrate how \Ac{ID2T} can be used to either reproduce or replicate results as well as to test the detection capabilities of an arbitrary setup of \Acp{NIDS}. In the first use case, we reproduce the network intrusion detection capabilities of \Acp{RNN} using different days of the MAWI dataset (in contrast to the days used in the original publication). Although the \Ac{PCAP} for training the \Ac{RNN} in the original publication and our use case come from MAWI, there are almost two years of difference between them. We argue that in the last two years the characteristics of network traffic in backbone networks have changed considerably~\cite{akamai2018}. In this context, \Ac{ID2T} was able to help us reproduce past results that also apply to modern network traffic without creating an updated dataset from scratch. The process of creating new synthetic attacks that \Ac{ID2T} can inject is considerably faster than publishing new datasets every time a novel attack or network configuration needs to be tested.

In the second use case scenario, we use \Ac{ID2T} to test two popular \Ac{NIDS}. \Ac{ID2T} is a tool that enabled the creation of network traffic that contained certain attacks that are easy to detect using the right signatures. We showed how a properly configured installation of the Bro and Suricata \Acp{NIDS} are able to detect expected attacks. \Ac{ID2T} can be used, therefore, to verify the correct operation of already deployed systems.


\section{Conclusion and Future Work}%
\label{sec:concl-future-work}

We developed \Ac{ID2T} to try to address the long standing issue of not having reliable and reproducible datasets in the \Acp{NIDS} community. To evaluate systems, researchers in this community have a tendency of using old, incomplete or private datasets that hamper their results. For example, it is a common occurrence to find new publications that use outdated datasets (such as the DARPA 1999 dataset). A system that demonstrates its capabilities on 19 year old network traffic cannot be trusted to work on modern traffic. The problem of acquiring reliable datasets lies in the difficulty of creating datasets that are useful and, at the same time, fulfill the conditions or constraints of the dataset's creators. Privacy, for example, is a major concern that ends up impacting a dataset because of the way data is anonymized or concealed.

With \Ac{ID2T}, we provide a tool that enables the creation of custom datasets while solving the problems of reproducibility and privacy preservation. Instead of providing a single static dataset, \Ac{ID2T} gives researchers the tools to inject synthetic attacks into background traffic they provide. The injected attacks replicate the characteristics of the background traffic to erase any trace that points towards the usage of \Ac{ID2T} and the existence of synthetic traffic. With such a tool, we envision researchers demonstrating the usage of their systems by specifying the type of background traffic used and the parameters used to create an attack. For example, as illustrated in our use-case section (see Section~\ref{sec:ident-attacks-using}), we use backbone network traffic to detect \Ac{DDoS} attacks and SMB probing scans.

One key functionality of \Ac{ID2T} is that of replicating the characteristics of the provided background traffic in the synthetic attacks it generates. Therefore, the properties of the background traffic impact the injected attacks. We developed the \Ac{TIDED} module of \Ac{ID2T} (see Section~\ref{sec:tided}) for the purpose of helping researchers to determine if the characteristics of the provided background traffic conform to their expectations. With such tests, for example, it is possible to determine if traffic that claims to come from a backbone network reflects the properties of such a large network. With \Ac{TIDED}, it is also possible to detect potential sources of defects that a dataset may have.

In future work, we wish to expand the capabilities of \Ac{ID2T} in two directions. First, we plan to enlarge the arsenal of injectable attacks to include recent popular attacks. Second, we would like to address the main limitation of \Ac{ID2T}. As it stands, \Ac{ID2T} has the limitation of not having a feedback loop between the generated attacks and the background traffic. In real scenarios, an attack affects the characteristics of the network onto which it was launched. A \Ac{DDoS} attack, for example, may cause a high number of TCP re-transmissions and lower average packet window sizes. The attacks generated by \Ac{ID2T} are limited to only replicating what is observed in the background traffic. It is also desirable, for the purpose of creating more realistic traffic, to model the interaction that an attack has on the background data. With such an interaction, attacks would also modify the background traffic to either add, modify or delete packets.


\bibliographystyle{ACM-Reference-Format}
\bibliography{Cybersecurity,manual}

\end{document}